\documentclass[aps,prx,twocolumn,superscriptaddress,floatfix,longbibliography]{revtex4-1}

\usepackage{graphicx, placeins, amsmath, braket, color, xcolor}
\usepackage[unicode=true,pdfusetitle,bookmarks=true,bookmarksnumbered=true,bookmarksopen=false, breaklinks=false,pdfborder={0 0 0},pdfborderstyle={},backref=false,colorlinks=false, pdftitle={Observation of a smooth polaron--molecule transition in a degenerate Fermi gas}]{hyperref}

\makeatletter
\renewcommand\frontmatter@abstractwidth{\dimexpr\textwidth-1in\relax}
\makeatother

\newcommand{\pv}{\mathbf{p}}
\newcommand{\kv}{\mathbf{k}}
\newcommand{\qv}{\mathbf{q}}
\newcommand{\lv}{\mathbf{l}}
\newcommand{\dpv}{\bar{\vecq}}
\newcommand{\pzero}{\mathbf{0}}
\newcommand{\veck}{\mathbf{k}}
\newcommand{\vecp}{\mathbf{p}}
\newcommand{\vecpp}{\mathbf{p}^{\prime}}
\newcommand{\vecq}{\mathbf{q}}
\newcommand{\qr}{\bar{\mathbf{q}}}
\newcommand{\bq}{\bar{q}}
\newcommand{\kq}{k_{\bar{q}}}
\newcommand{\kcm}{k_{\mathrm{cm}}}
\newcommand{\krel}{k_{\mathrm{rel}}}
\newcommand{\Ep}{\epsilon_{\mathrm{pol}}}
\newcommand{\Em}{\epsilon_{\mathrm{mol}}}
\newcommand{\EF}{\varepsilon_F}
\newcommand{\kfainv}{\left(k_Fa\right)^{-1}}

\begin{document}
\title{Observation of a smooth polaron--molecule transition in a degenerate Fermi gas}

\author{Gal Ness}
\affiliation{Physics Department, Technion -- Israel Institute of Technology, Haifa 32000, Israel}
\author{Constantine Shkedrov}
\affiliation{Physics Department, Technion -- Israel Institute of Technology, Haifa 32000, Israel}
\author{Yanay Florshaim}
\affiliation{Physics Department, Technion -- Israel Institute of Technology, Haifa 32000, Israel}

\author{\\Oriana K. Diessel}
\affiliation{Max-Planck-Institute of Quantum Optics, Hans-Kopfermann-Strasse. 1, 85748 Garching, Germany}
\affiliation{Munich Center for Quantum Science and Technology (MCQST), Schellingstr. 4, 80799 M\"unchen, Germany}

\author{Jonas von Milczewski} 
\affiliation{Max-Planck-Institute of Quantum Optics, Hans-Kopfermann-Strasse. 1, 85748 Garching, Germany}
\affiliation{Munich Center for Quantum Science and Technology (MCQST), Schellingstr. 4, 80799 M\"unchen, Germany}

\author{Richard Schmidt}
\affiliation{Max-Planck-Institute of Quantum Optics, Hans-Kopfermann-Strasse. 1, 85748 Garching, Germany}
\affiliation{Munich Center for Quantum Science and Technology (MCQST), Schellingstr. 4, 80799 M\"unchen, Germany}

\author{Yoav Sagi}
\email[Electronic address: ]{yoavsagi@technion.ac.il}
\affiliation{Physics Department, Technion -- Israel Institute of Technology, Haifa 32000, Israel}

\date{\today}

\begin{abstract}
Understanding the behavior of an impurity strongly interacting with a Fermi sea is a long-standing challenge in many-body physics. When the interactions are short-ranged, two vastly different ground states exist: a polaron quasiparticle and a molecule dressed by the majority atoms. In the single-impurity limit, it is predicted that at a critical interaction strength, a first-order transition occurs between these two states. Experiments, however, are always conducted in the finite temperature and impurity density regime. The fate of the polaron-to-molecule transition under these conditions, where the statistics of quantum impurities and thermal effects become relevant, is still unknown. Here, we address this question experimentally and theoretically. Our experiments are performed with a spin-imbalanced ultracold Fermi gas with tunable interactions. Utilizing a novel Raman spectroscopy combined with a high-sensitivity fluorescence detection technique, we isolate the quasiparticle contribution and extract the polaron energy, spectral weight, and the contact parameter. As the interaction strength is increased, we observe a continuous variation of all observables, in particular a smooth reduction of the quasiparticle weight as it goes to zero beyond the transition point. Our observation is in good agreement with a theoretical model where polaron and molecule quasiparticle states are thermally occupied according to their quantum statistics. At the experimental conditions, polaron states are hence populated even at interactions where the molecule is the ground state and vice versa. The emerging physical picture is thus that of a smooth transition between polarons and molecules and a coexistence of both in the region around the expected transition. Our findings establish Raman spectroscopy as a powerful experimental tool for probing the physics of mobile quantum impurities and shed new light on the competition between emerging fermionic and bosonic quasiparticles in non-Fermi-liquid phases.
\end{abstract}

\maketitle

\section{Introduction}
In order to understand the motion of an electron through an ionic lattice, Landau suggested treating the electron and the phonons that accompany its movement as a new quasiparticle named `polaron' \cite{Landau_first_polaron_paper}. The concept of the polaron was later found to be applicable in many other systems, including semiconductors \cite{Lindemann1983}, high-temperature superconductors \cite{Mott1993}, alkali halide insulators \cite{Popp1972}, and transition metal oxides \cite{Moser2013}. In such systems, polarons appear as weakly- or strongly-coupled quasiparticles that are classified as large or small, depending on the size of the distortion they generate in the underlying crystalline structure of the material \cite{Emin2012}. Understanding the properties of polarons coupled to a bosonic bath is still an ongoing effort in areas ranging from solid-state physics \cite{Alexandrov2010,Sio2019a} and ultracold atoms \cite{hvk16,jws16,Yan2019}, to quantum chemistry \cite{lemeshko2016}. 

The concept of polarons becomes also a powerful tool for our understanding of the properties of quantum impurities interacting with a fermionic environment. In this context, applications range from ions in liquid $^3$He \cite{kondo1983theory}, and mixtures of cold atomic gases  \cite{Massignan2014} to excitons interacting with electrons in atomically thin semiconductors \cite{Sidler2016,Efimkin2017,Fey2020}. Strikingly, in fermionic systems an infinite number of low-energy excitations leads to the Anderson orthogonality catastrophe for immobile impurities and a complete loss of quasiparticle behavior \cite{anderson1967}. In contrast, for mobile impurities, the energy cost related to the impurity recoil stabilizes the formation of Fermi polarons with well-defined quasiparticle properties \cite{rosch1999quantum}. 

One of the simplest scenarios in which Fermi polarons naturally emerge is in ultracold gases, where a small number of (mobile) spin impurities can be immersed in a system of free fermions of the opposite spin. Such ultracold, spin-imbalanced systems are ideally suited to explore polaron physics \cite{Chevy2010,Massignan2014,Schmidt2018} owing to their extremely long spin-relaxation times and tunability of
the $s$-wave scattering length, $a$, between the impurity and the majority atoms via Feshbach resonances \cite{RevModPhys.82.1225}. 

Initial experiments with spin-imbalanced Fermi gases in harmonic confinement revealed phase separation into three regions at unitary interactions ($a\rightarrow \infty$). It was observed that phases arrange according to the varying local density, with an inner core of a spin-balanced superfluid being separated from a second shell of a partially polarized normal gas, and a third shell of a fully polarized gas \cite{Zwierlein27012006,Partridge27012006,PhysRevLett.97.030401}. It was Chevy who first pointed out \cite{Chevy2006} that the radius between the outer and intermediate shells is related to the solution of the Fermi polaron problem \cite{Prokofev2008}. For weak attractive interactions ($a<0$), the ground state is a long-lived quasiparticle dressed by the majority particles, forming the attractive polaron. Beyond the Feshbach resonance, at $a>0$, it was found that a metastable polaronic state also exists energetically far up in the excitation spectrum \cite{Cui2010,Schmidt2011,massignan2011repulsive,Kohstall2012,Koschorreck2012,Schmidt2012b,Ngampruetikorn2012,Oppong2019}. This so-called repulsive polaron becomes, however, progressively unstable towards unitary interactions. 

Fermi polarons have well-defined momenta with a narrow dispersion relation that is described by a renormalized effective mass \cite{Combescot2007,Schmidt2011}. The attractive polaron persists as the ground state even as the interactions increase towards unitarity. However, for still stronger interactions, the system favors a molecular ground state dressed by the majority fermions \cite{Prokofev2008,Mora2009,Punk2009} (see Fig.~\ref{Fig:RamanSpectraTheory2}(a) below). It was predicted that the energies of the polaron and molecular states cross around $\kfainv_c\approx 0.9$ ---with $k_F$ being the Fermi wave vector of the majority--- leading to a sharp, first-order transition between the two ground states \cite{Prokofev2008,Prokofev2008a,Punk2009,Mora2009,Combescot2009,Schmidt2011}. Contrasting claims for a smooth crossover were also put forward \cite{Edwards2013,Chen2016,Tajima2018,Cui2020}.

The Fermi polaron problem represents the limiting case of a spin-imbalanced Fermi gas. Therefore, the nature of the polaron-to-molecule transition has profound theoretical implications for the phase diagram of the spin-imbalanced BEC-BCS crossover \cite{nikolic2007renormalization,Punk2009,Zwerger_book,zwerger2016strongly,frank2018}. While at zero temperature, the polaron-to-molecule transition was predicted to be pre-empted by phase separation between the superfluid and the normal phases \cite{PhysRevLett.100.030401}, at finite temperature, increased thermal fluctuations are expected to suppress the superfluid and restore the polaron-to-molecule transition. Experimentally, the Fermi polaron was initially studied by radio-frequency (rf) spectroscopy, where the attractive polaron was identified by a narrow peak appearing exclusively in the minority spectrum \cite{PhysRevLett.102.230402}. The spectral weight of this peak was interpreted as the quasiparticle residue (or weight) $Z$, which quantifies how similar the polaron remains to the non-interacting impurity particle. Accordingly, it is determined by the overlap between the polaron wavefunction and its non-interacting impurity state. When $Z$ is zero, the quasiparticle description is no longer valid. In the experiment, $Z$ was observed to continuously decrease and vanish above a certain interaction strength in contrast to theoretical predictions based on the Chevy Ansatz wavefunction \cite{Punk2009}. 

A different approach to measure $Z$ was employed in Ref.~\cite{Kohstall2012}. Here, $Z$ was determined from coherent oscillations between the polaron and a non-interacting impurity state. In this approach, the coherent oscillations address the polaronic state even when the attractive polaron is an excited state above the molecular ground state. This made it possible to measure the weight $Z$  across the full polaron-to-molecule transition, and it was found that $Z$ indeed does not vanish beyond $\kfainv_c$. Further properties of  attractive and repulsive Fermi polarons were also determined, including their effective mass \cite{PhysRevLett.103.170402,Scazza2017}, energy \cite{PhysRevLett.102.230402,Koschorreck2012,Kohstall2012,Scazza2017}, thermodynamics \cite{Zhenjie2019}, equation of state \cite{Navon07052010}, and formation dynamics \cite{Cetina2016}. However, despite these tremendous efforts, the fate of the polaron-to-molecule transition at realistic conditions, namely finite temperature and impurity concentration, remains unknown. Importantly, the key question whether the first-order polaron-to-molecule transition prevails at finite impurity density, and separates sharply a phase of polarons from a gas of dressed molecular quasiparticles, is still open. 

In this work, we address this question both theoretically and experimentally. Our experiments are performed with a spin-imbalanced, ultracold Fermi gas in the BEC-BCS crossover regime \cite{Zwerger_book,RevModPhys.82.1225}. To gain detailed insight into the behavior of the quasiparticles, we employ a novel spectroscopic method based on a two-photon Raman transition. Raman spectroscopy allows us to clearly identify the coherent response of the polarons, and determine some of its key properties, including its energy and spectral weight. We compare our results with a theoretical model that takes into account the thermal occupation of polarons and molecules at finite momenta. Both our theoretical model and the measurements consistently show that a finite impurity concentration and temperature have a striking effect on the transition: they smooth it and lead to a regime where polarons and molecules coexist.

After describing the experimental setup in Section~\ref{Sec:Experiment}, we briefly introduce our theoretical model in Section~\ref{sect:FermiPolaronModel} and present the calculated Raman spectra. Based on this, in Section~\ref{sec:analysis} we develop a fitting routine for the experimental spectra which allows us to extract physical quantities, such as the polaron energy, the quasiparticle spectral weight, and the contact parameter. The experimental results are presented and discussed in Section~\ref{Sec:ExperimentalResults}, while in  Section~\ref{sect:Theory2} we give a detailed theoretical derivation of our model. In particular, we show how the quantities accessible in the experiment are computed. Finally, in Section~\ref{Discussion}, we summarize our results, discuss their implications for the many-body physics of cold Fermi gases and outline directions for future work.

\begin{figure}
	\centering
	\includegraphics[width=1\linewidth]{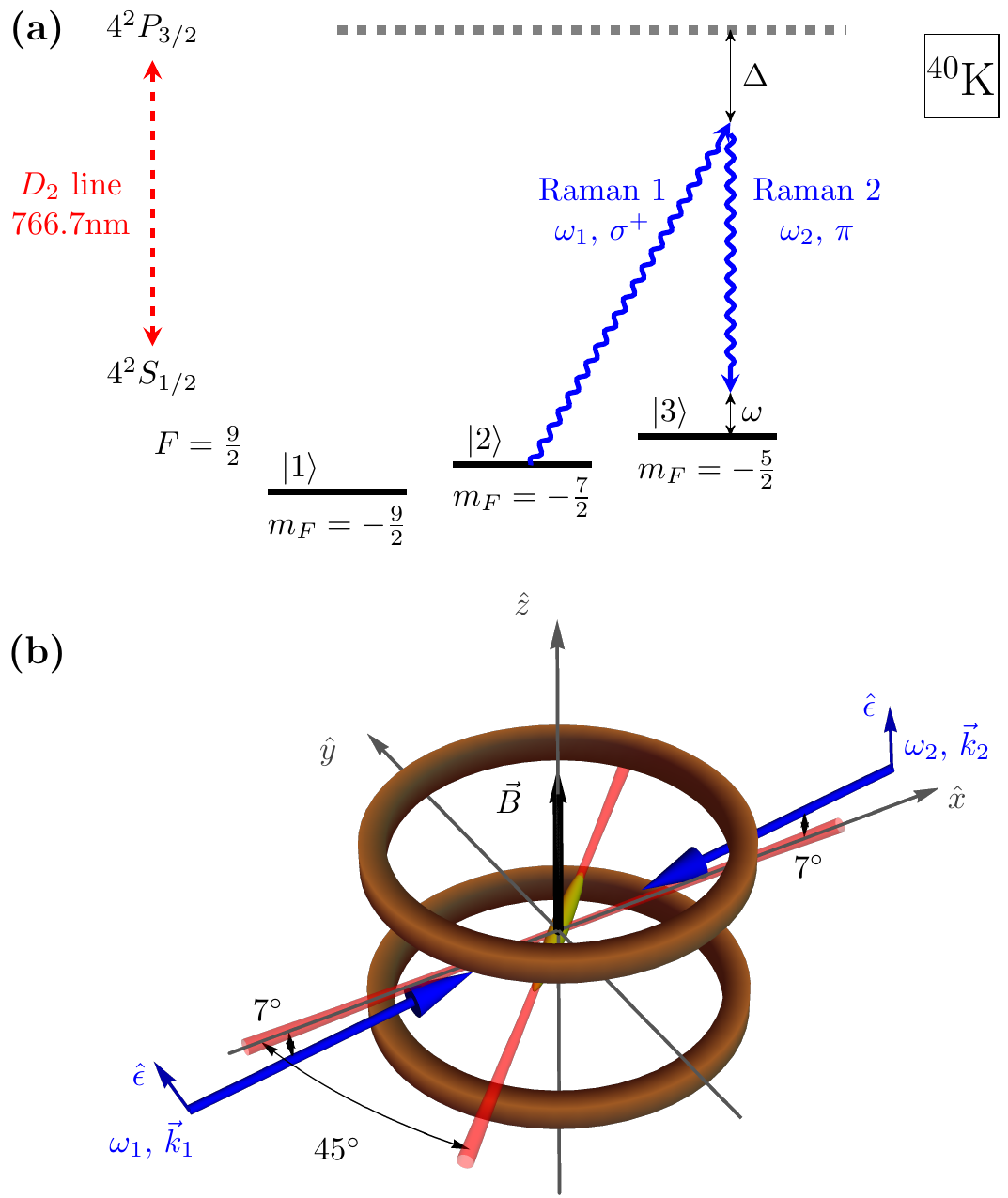}
	\caption{\textbf{Experimental setup.} \textbf{(a)} Level diagram of the relevant states in $^{40}$K. The majority of atoms occupy state $\left|1\right\rangle=\left|F=9/2,m_F=-9/2\right\rangle$ in the $4^2\mathrm{S}_{1/2}$ manifold, and their interaction with the minority atoms in state $\left|2\right\rangle=\left|F=9/2,m_F=-7/2\right\rangle$ can be tuned in the vicinity of the Feshbach resonance at $B\approx202.14$G \cite{Shkedrov2018}. The counter-propagating Raman beams (wiggly blue lines) are pulsed for $500\mu$s and couple atoms in state $\left|2\right\rangle$ to state $\left|3\right\rangle=\left|F=9/2,m_F=-5/2\right\rangle$, which is initially unoccupied. Afterwards, we detect the number of atoms in state $\left|3\right\rangle$ \cite{Shkedrov2018}. The single-photon Raman detuning is $\Delta=-2\pi\times 54.78(8)$GHz relative to the $D_2$ transition. \textbf{(b)} 3D sketch of the beam configuration in the experiment. Two Raman beams with orthogonal polarization (blue lines with arrowheads) overlap the atomic cloud (yellow), which is being held in an elongated crossed-beams optical dipole trap (red lines). The optical trap oscillation frequencies are $\omega_\mathrm{radial}=2\pi\times 238(3)$Hz and $\omega_\mathrm{axial}=2\pi\times 27(2)$Hz, in the radial and axial directions, respectively. The gravitational acceleration direction is $-\hat{z}$.}
	\label{fig:sketch1}
\end{figure}

\section{The experiments}
\label{Sec:Experiment}

\begin{figure}[b]
    \centering
	\includegraphics{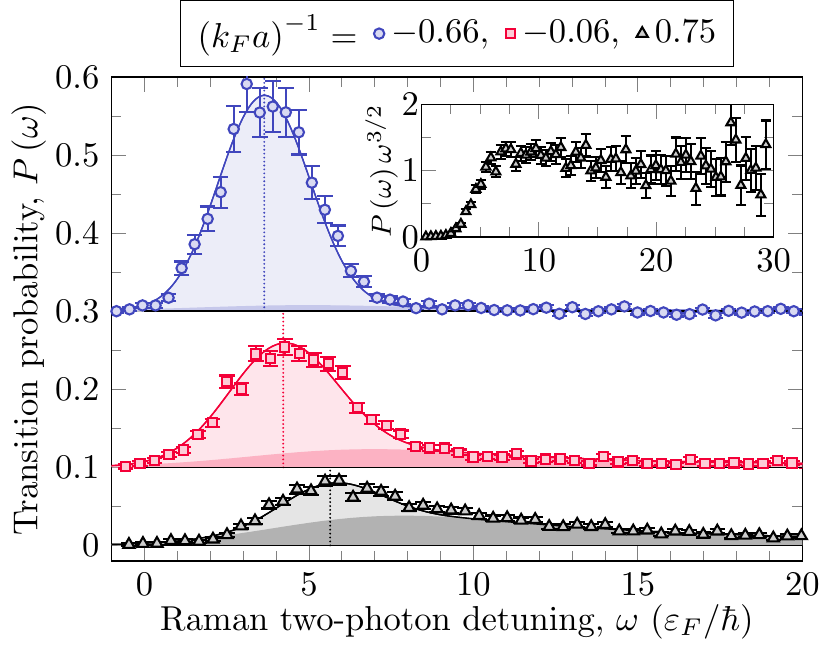}
	\caption{\textbf{Raman spectra of an imbalanced Fermi gas in the BEC-BCS crossover.} The Raman transition probes the minority atoms, which represent around $4\%$ of the whole cloud. The data is fit with the function given in Eq.\,\eqref{Eq_trans_prob_pairs_and_QP} (solid lines). The dark shaded area under the graph is the combined contribution from  molecules and the incoherent part of the polaron wavefunction, $P_{\mathrm{bg}}\left(\omega;T_{\mathrm{bg}},E_b\right)$, while the light shaded area is the coherent polaron contribution, $P_{\mathrm{coh}}\left(\omega;T_p,\Ep^0,m^\ast\right)$. We extract the polaron energy, $\Ep^0$, from the peak position (dotted vertical lines), shifted by the  recoil energy due to the finite Raman momentum transfer of $\bq\approx1.9\,k_F$. The second and third graphs from the bottom are vertically shifted for clarity by $0.1$ and $0.3$, respectively. In the inset, we depict the data at $\kfainv=0.75$ multiplied by $\omega^{3/2}$ to demonstrate the power-law scaling of the high-frequency tail. Each point is an average of three experiments, and the area below each curve is normalized to unity. Since the measurement is done up to some maximal frequency, to properly normalize the data we must account for the missing spectral weight in the unmeasured tail. This is done by adding to the normalization the integral of $\omega^{-3/2}$ up to an energy cutoff of $\hbar^2/m r_0^2$, where $r_0\approx181\,a_0$ is the effective range of the interparticle potential \cite{RevModPhys.82.1225}.
	}
	\label{fig:raman_data_for_polarons}
\end{figure}

The experiments are performed with a harmonically-trapped ultracold gas of $^{40}$K atoms. The system is initially prepared in an incoherent mixture of the two lowest Zeeman states denoted by $\left|1\right\rangle$ and $\left|2\right\rangle$ (see Fig.~\ref{fig:sketch1}), with the majority of atoms being in state $\left|1\right\rangle$. The cooling sequence is similar to the one described in Ref.~\cite{Shkedrov2018}, here modified to produce a spin-polarized ensemble of $\sim 100,000$ atoms in the state $\left|1\right\rangle$ at a temperature of $T\approx0.2\,T_F$, where $T_F$ is the Fermi temperature. This is achieved by terminating the evaporation cooling at a magnetic field of $201.75$G, below the Feshbach resonance ($B_0\approx202.14$G \cite{Shkedrov2018}), where three-body processes remove all the atoms in state $\left|2\right\rangle$. The magnetic field is then ramped adiabatically to the BCS side of the Feshbach resonance ($204.5$G), where the interaction between the states $\left|1\right\rangle$ and $\left|2\right\rangle$, parametrized by the $s$-wave scattering length $a$, is weak. To introduce the impurities, we use a short (few microseconds) rf pulse that transfers a very small fraction of the atoms from state $\left|1\right\rangle$ to  $\left|2\right\rangle$. This is followed by a hold time of $100$ms during which  the two states fully decohere. Finally, the magnetic field is ramped adiabatically to its final value where we wait another $3.3$ms before pulsing  the Raman beams for $500\mu$s.

The minority concentration $x$, can be defined globally by $x=\left.N_I\right/N$, with $N_I$ ($N$) being the total number of impurity (bath) atoms, or alternatively by averaging over its local value $\left\langle x\right\rangle$ in the harmonic trap (see Appendix~\ref{appendix_local_density_approximation}). Since the local density $n_I\left(\mathbf{r}\right)$ depends on $k_Fa$, $\left\langle x\right\rangle$ changes even when $x$ is kept constant. To ensure there are no systematic deviations in the experiments due to this effect, we have repeated the measurements twice: once keeping $x$ at approximately $0.04$, which gives $\left\langle x\right\rangle\approx 0.23$ at $\kfainv=0.85$, and a second time maintaining the same value of $\left\langle x\right\rangle$ by varying $x$. Since we did not observe any significant difference between the two datasets, in what follows we shall present their results together.

The main innovation in our experiments is the use of Raman spectroscopy. In conventional rf spectroscopy, the photon momentum is negligible, and the atomic momentum is essentially unchanged in the transition. As a result, the transition probability depends only weakly on the atom velocity and the maximal signal is attained for atoms that are not at rest. In particular, at finite impurity density, the measured peak depends on temperature \cite{Yan2019} (see Appendix~\ref{app:Ep_shift}). In contrast, in a Raman process, the momentum change is significant compared to the atomic momentum, and consequently, the transition rate depends on the atomic velocity. As we show further below, the Raman spectrum reflects the projection of the polaron momentum distribution along the two-photon Raman wavevector which, due to the symmetry of the resulting spectrum, allows us to uniquely identify the coherent contribution of the polarons.

As illustrated in Fig.~\ref{fig:sketch1}, in our setup  two Raman beams couple atoms in the minority state $\left|2\right\rangle$ to a third state $\left|3\right\rangle$, which is initially unoccupied. The beam parameters are the same as described in Ref.~\cite{Shkedrov2020}. We denote their frequencies by $\omega_1$ and $\omega_2$ and their wavevectors by $\mathbf{k}_1$ and $\mathbf{k}_2$. The measurement is performed by recording the number of atoms transferred to the state $\left|3\right\rangle$ versus the two-photon detuning, $\omega=\omega_1-\omega_2-E_0/\hbar$, where $E_0$ is the bare transition energy between states $\left|2\right\rangle$ and $\left|3\right\rangle$. To achieve the utmost sensitivity, we measure the atoms using a high-sensitivity fluorescence detection scheme we have recently developed \cite{Shkedrov2018,Shkedrov2020}.

In Fig.~\ref{fig:raman_data_for_polarons} we depict three representative experimental datasets taken on the BCS side ($\kfainv=-0.66$, blue circles), unitarity ($\kfainv=-0.06$, red squares) and on the BEC side ($\kfainv=0.75$, black triangles). The spectrum is symmetric on the BCS side, but becomes asymmetric towards unitarity with a tail at high frequencies that grows to be the dominant spectral feature on the BEC side. As we will see below, the symmetric part of the spectrum is associated with the coherent response of polarons, while the asymmetric contribution is due to the polaron incoherent part as well as molecules. The  peaked response of the quasiparticles arises since for the coherent contribution of the polarons the Raman transition rate is proportional to the one-dimensional momentum distribution, which is symmetric at equilibrium ($k\rightarrow-k$ invariant, neglecting effective mass variation). Molecules, on the other hand, are dissociated by the Raman process. Similar to the incoherent contribution from polarons, this opens another degree of freedom, namely the relative motions of the two atoms, which gives rise to the asymmetric energy tail in the Raman spectra. As can be seen in the inset of Fig.~\ref{fig:raman_data_for_polarons}, this tail features a power-law scaling of $\omega^{-3/2}$ at large $\omega$, as expected from the Tan contact relations \cite{Tan08,Zwerger_book,Nishida}.

\section{Fermi polaron model}
\label{sect:FermiPolaronModel}

\begin{figure*}
	\centering
	\includegraphics{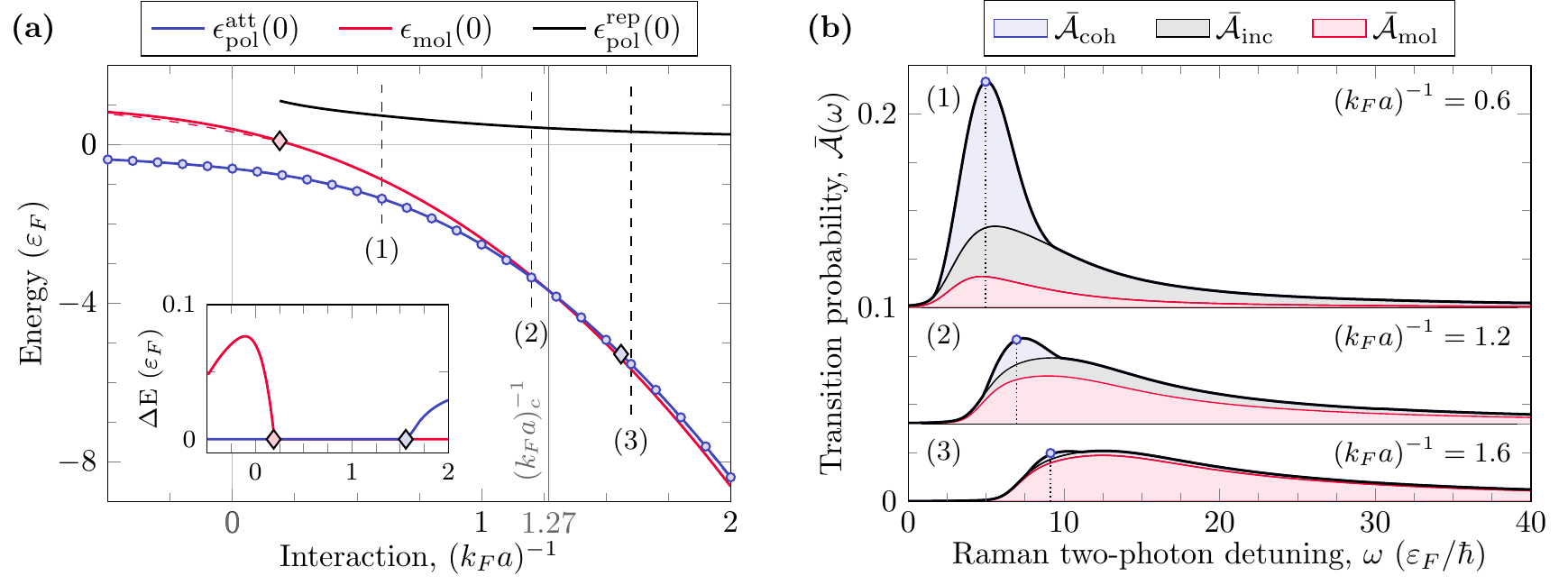}
    \caption{\textbf{Theoretical Raman spectra and quasiparticle energies.} \textbf{(a)}  Quasiparticle energies of different polaron and molecule states. The solid lines display the  energy of the molecule (red), the attractive (blue) and the repulsive (black) polaron at zero momentum as a function of the interaction strength, calculated from the variational states in  Eqs.~(\ref{Eq:WaveFunctionPolaron}) and (\ref{Eq:WaveFunctionMolecule}). The crossing of the polaron and molecule energies at $\kfainv_c\approx 1.27$ results in a first-order transition from a polaronic to a molecular ground state. The blue circles show the peak position of the coherent part of the Raman spectrum. The comparison with the solid blue line makes it evident that the zero-momentum polaron energy can reliably be extracted using this approach. For the excited polaron and molecular states, the energy minimum does not always appear at zero momentum. We show the corresponding lowest energies of the attractive polaron and the molecule at finite momentum as dashed lines. Since the energies are almost degenerate, we indicate the interaction strengths at which the energy minimum switches to finite momentum by a diamond symbol. The inset shows the difference between the zero-momentum energy and the minimal energy for both polaron (blue) and molecule (red) branches. 
    \textbf{(b)} Many-body Raman spectra of the imbalanced Fermi gas for three interaction strengths, calculated from an occupation average for fixed impurity concentration $N_I=0.15\,N$ and temperature $T=0.2\,T_F$.
    The blue-, red- and grey-shaded regions represent the coherent and incoherent polaronic contributions from (\ref{Eq:WaveFunctionPolaron}) as well as the molecular contributions from (\ref{Eq:WaveFunctionMolecule}), respectively. The coherent part yields an almost symmetric lineshape that is peaked at the polaron energy shifted by the constant Raman two-photon recoil energy (vertical dotted line),
    while the incoherent and molecular parts lead to an asymmetric continuum. For clarity, the second and third graphs from the bottom are shifted by $0.04$ and $0.1$, respectively.
  }
\label{Fig:RamanSpectraTheory2}
\end{figure*}

In order to analyze the experimentally measured Raman spectra in terms of the physics of polarons and molecules, we consider the Hamiltonian $\mathcal{H}$ describing a system of fermionic impurities immersed in a homogeneous fermionic bath,
\begin{align}
\mathcal{H}&=\sum_{\pv}\varepsilon_\pv^{\phantom{\dagger}} c_\pv^{\dagger}c_\pv^{\phantom{\dagger}}+\sum_\pv^{\phantom{\dagger}}\varepsilon_\pv^{\phantom{\dagger}}d^{\dagger}_\pv d_\pv^{\phantom{\dagger}}
\nonumber\\
&\qquad+\frac{U}{V}\sum_{\vecp,\vecpp,\vecq}c_{\vecp+\vecq}^{\dagger} c_\vecp^{\phantom{\dagger}} d^\dagger_{\vecpp-\vecq}d_{\vecpp}^{\phantom{\dagger}}\;.
\label{Eq:Hamiltonian}
\end{align}
Here $V$ is the system volume, and the operators $c^{\dagger}_\pv$ and $ d^{\dagger}_\pv$ denote fermionic creation operators of bath ($\left|1\right\rangle$) and impurity ($\left|2\right\rangle$) particles, respectively (see Fig.~\ref{fig:sketch1}). Both species have the same mass $m$ and their free dispersion relation is given by $\varepsilon_\pv=\pv^2/2m$ (unless indicated otherwise, we work in units of $\hbar=k_B=1$). 
The interaction between impurity and bath particles is modeled by the contact interaction in the last term of Eq.\,\eqref{Eq:Hamiltonian}, which is an excellent approximation for open-channel dominated Feshbach resonances as employed in our experiment \cite{RevModPhys.82.1225}. Its strength $U$ is related to the $s$-wave scattering length $a$ by the Lippmann--Schwinger equation $U^{-1}=m/4 \pi a-V^{-1} \sum_\kv 1/2 \varepsilon_{\kv}$. 

\textbf{\textbf{Polarons and molecules.}---} The physics of polarons and molecules can be qualitatively understood in terms of two sets of variational wavefunctions that approximate the exact eigenstates of the Hamiltonian in Eq.\,\eqref{Eq:Hamiltonian}. On the one hand, the formation of polarons is well described by an Ansatz that systematically expands the many-body wavefunction in terms of particle-hole excitations of the Fermi sea. It was found \cite{Prokofev2008,Prokofev2008a,PhysRevLett.101.050404,Liu2019} that accounting only for a single such excitation in form of the Chevy Ansatz \cite{Chevy2006,Trefzger2012}, 
\begin{equation}
\ket{\psi_P^{\pv}}= \alpha_0^{\pv} d_\pv^{\dagger}\ket{\text{FS}_N}+\sum_{\kv,\qv}{}^{'}\alpha_{\kv,\qv}^{\pv}d^{\dagger}_{\pv+\qv-\kv}c^{\dagger}_\kv c_{\qv}^{\phantom{\dagger}}\ket{\text{FS}_N}
\label{Eq:WaveFunctionPolaron}
\end{equation}
already yields a remarkably good approximation for the attractive polaron at momentum $\vecp$. Here, $\alpha_0^{\pv}$ and $\alpha_{\kv,\qv}^{\pv}$ are variational parameters, and primed sums indicate that the summation is taken over momenta fulfilling $|\kv|>k_F$ and $|\qv|<k_F$. The state $\ket{\text{FS}_N}$ denotes the zero temperature Fermi sea of $N$ majority particles (the Fermi wave vector $k_F$, and energy $\varepsilon_F$ refer to the majority ensemble). Crucially, the first term describes the so-called coherent part of the polaron wavefunction. At low polaron momenta it determines the quasiparticle weight $Z_\vecp = |\alpha_0^\vecp|^2$. It results in a coherent quasiparticle peak in the spectroscopic measurements, while the second term, describing the entanglement of the impurity with bath degrees of freedom, leads to an incoherent background of asymmetric shape. Importantly, while the polaron quasiparticle weight $Z_\vecp$ is finite for all interactions, the polaron becomes an excited state beyond a critical interaction strength. It is thus not occupied and hence ---at the polaron-to-molecule transition--- a jump in the spectral response is expected.

On the other hand, to lowest order, molecular states at momentum $\vecp$ can be described by an Ansatz with variational parameters $\beta_{\kv}^{\pv}$  of the form \cite{Punk2009,Mora2009,Trefzger2012}
\begin{equation}
\ket{\psi_M^\pv}=\sum_{\kv}{}^{'} \beta_{\kv}^{\pv}c^{\dagger}_{-\kv}d^{\dagger}_{\kv+\pv}\ket{\text{FS}_{N-1}}\;.
\label{Eq:WaveFunctionMolecule}
\end{equation}
Here, a fermion is removed from the Fermi surface and is paired with the impurity particle. Both expressions in Eqs.\,\eqref{Eq:WaveFunctionPolaron} and \eqref{Eq:WaveFunctionMolecule} can be systematically improved in their accuracy by entangling a larger number of particle-hole excitations of the Fermi sea with the quantum impurity. 

The minimization of the energy functional $\braket{\psi_{P/M}^\pv|\mathcal{H}-E|\psi_{P/M}^\pv}$ with respect to the variational parameters allows one to determine the renormalized dispersion relations $\Ep(\pv)$ and $\Em(\pv)$ of the polaron and the molecule, respectively. The wavefunctions in Eqs.\,\eqref{Eq:WaveFunctionPolaron} and \eqref{Eq:WaveFunctionMolecule} predict that the energies of the polaron and molecule cross at an interaction strength $\kfainv_c\approx 1.27$ (see Fig.~\ref{Fig:RamanSpectraTheory2}(a)). The deviation from the state-of-the-art prediction of $\kfainv_c\approx 0.90(4)$, obtained from diagrammatic Monte Carlo (diagMC) calculations \cite{Prokofev2008,Prokofev2008a,Werner2020}, is mostly due to the simple approximation taken for the molecular Ansatz that neglects particle-hole dressing of the molecular state \cite{Punk2009}.

\textbf{\textbf{Raman spectroscopy.}---}
For a single impurity, the Raman transition rate $\Gamma(\omega,i)= 2 \pi \Omega_e^2 \mathcal{A}(\omega,i)$ is given by Fermi's golden rule as
\begin{align}
\mathcal{A}(\omega ,i)=\sum_f \big|\braket{ f|\hat {V}_R|i}\big|^2\delta\left[\omega-\left(E_f-E_i\right)\right]\;,
\label{Eq:FermisGoldenRule}
\end{align}
where $\mathcal{A}(\omega,i)$ denotes a Raman spectrum and $\Omega_e$ is the effective Rabi frequency. Here, the impurity resides in an initial state $|i\rangle$ that is characterized by a conserved momentum $\vecp$ and may be either polaronic or molecular; i.e. $\ket{i}=\{\ket{\psi^\vecp_P},\ket{\psi^\vecp_M}\}$ (such as, e.g., approximately given by Eqs.\,\eqref{Eq:WaveFunctionPolaron} and \eqref{Eq:WaveFunctionMolecule}). The corresponding energies are $E_i=\left\{\Ep(\vecp)\right.$, $\left.\Em(\pv)\right\}$, respectively.

The  operator $ \hat V_R=\sum_\vecp ( f^\dagger_{\vecp+\bar\vecq} d_\vecp^{\phantom{\dagger}} + h.c.) $ describes the transition with relative two-photon momentum $\bar \vecq = \veck_2-\veck_1$ from an interacting impurity state ($\ket{2}$ in the experiment) at momentum $\vecp$, to a hyperfine state (created by $ f_\veck^\dagger$; state $\ket{3}$ in the experiment) that is decoupled from the fermionic environment and  governed by the Hamiltonian $\mathcal{H}_f=\sum_{\pv}\varepsilon_\pv^{\phantom{\dagger}}(c_\pv^{\dagger}c_\pv^{\phantom{\dagger}}+f^{\dagger}_\pv f_\pv^{\phantom{\dagger}})$.  The final states $\ket{f}$ of energy $E_f$ are thus given by non-interacting continuum states such as $f_\vecp^{\dagger}\ket{\text{FS}_N}$ or $f^{\dagger}_{\vecp}c^{\dagger}_\veck c_{\vecq}^{\phantom{\dagger}}\ket{\text{FS}_N}$. 

In experiments, the number of impurities $N_I=n_I V$ is finite.
Treating this case theoretically simplifies at sufficiently low impurity concentration, where one may assume that impurities occupy polaronic and molecular eigenstates of the form of Eqs.\,\eqref{Eq:WaveFunctionPolaron} and \eqref{Eq:WaveFunctionMolecule}, respectively. Importantly, this allows polaron states to be occupied at finite momentum even in the regime where the molecule is the ground state, and vice versa, for the molecule before the polaron-to-molecule transition (for a detailed discussion see Section\,\ref{sect:Theory2}). 
Each of these occupied states contributes to the total normalized Raman signal which is thus obtained by averaging over occupation numbers of polarons and molecules with associated Fermi and Bose distribution functions 
\begin{align}
    \bar{\mathcal{A}}(\omega)=&\frac{1}{N_I}\sum_\pv \mathcal{A}_\mathrm{pol}(\omega,\vecp) \cdot n_F\left[\Ep(\pv) \right]\nonumber\\
   &+\frac{1}{N_I}\sum_\pv \mathcal{A}_\mathrm{mol}(\omega,\vecp)\cdot n_B\left[\Em(\pv) \right]\;.
    \label{Eq:FullRamanSpectrum}
\end{align}
Here, $\mathcal{A}_\mathrm{pol}(\omega,\vecp)\equiv \mathcal{A}(\omega ,i=\ket{\psi_P^\vecp})$ and analogously for $\mathcal{A}_\mathrm{mol}(\omega,\vecp)$. Similar to the single impurity case, the full many-impurity Raman signal is connected to the Raman rate by $\bar{\Gamma}(\omega,i)= 2 \pi \Omega_e^2 N_{I} \bar{\mathcal{A}}(\omega,i)$.

\textbf{Model prediction.---} Theoretical Raman spectra based on wavefunctions Eqs.\,\eqref{Eq:WaveFunctionPolaron} and \eqref{Eq:WaveFunctionMolecule} are shown in Fig.~\ref{Fig:RamanSpectraTheory2}(b) for three interaction strengths before and after the polaron-to-molecule transition. Overall, the calculated spectra are qualitatively very similar to the measured ones shown in Fig.~\ref{fig:raman_data_for_polarons}. Each spectrum is composed of a polaronic and a molecular contribution. As can be seen, the polaronic contribution is separated into a coherent (blue shading) and an incoherent part (black shading). The former arises from the coherent part of the polaron wavefunction proportional to $\left|\alpha_0^{\vecp}\right|^2$,  and thus gives access to the quasiparticle weight $Z_{\vecp}$ of polarons. The incoherent polaron contribution, in turn, is due to the second term in Eq.\,\eqref{Eq:WaveFunctionPolaron}, leading to a highly asymmetric lineshape. The molecular contribution features a similar asymmetric lineshape as Raman transitions dissociate molecules into relative momentum states described by the variational parameters $\beta_\kv^\pv$ in Eq.\,\eqref{Eq:WaveFunctionMolecule}. Both the incoherent polaron and molecular contributions involve superposition states at large momenta (and thus probing short-distance physics); hence, their spectra  feature the characteristic high-frequency $\sim \omega^{-3/2}$ tails which were experimentally observed.

Our calculations show that Raman spectroscopy provides a tool for a clear dissection of the polaron state. This is due to the fact that ---in contrast to rf spectroscopy---  there is a qualitative difference between an almost symmetric response arising from the coherent polaron contribution (blue shading in Fig.~\ref{Fig:RamanSpectraTheory2}(b)) and a combined asymmetric response stemming from the incoherent part of the polaron wavefunction and from molecules (black and red shading, respectively). It is this clear distinction between the coherent and incoherent/molecular response that enables us to experimentally extract polaron quasiparticle properties with high detail using Raman spectroscopy.

\section{Quantitative analysis of experimental Raman spectra\label{sec:analysis}}

In order to quantitatively analyze the experimental data, we devise a fit model for the lineshape of the Raman transition amplitude. Here we make use of the fact that, although the theoretical spectra are approximations, they reveal general characteristics of the response, namely that it is composed of two main contributions:
\begin{enumerate}
    \item \textbf{Coherent polaron peak $P_{\mathrm{coh}}(\omega)$:} a roughly symmetric peak due to the coherent part of the polaron. It contains information about quasiparticle properties such as the polaron energy $\Ep$, the spectral weight $Z$, and the effective mass $m^\ast$. In particular, we find that the coherent part spectrum is proportional to the polaronic momentum distribution, and its peak position gives the polaron energy at zero momentum $\Ep^0$ (plus the recoil energy from the two-photon Raman transition, see Eq.\,\eqref{Eq_omega_0_of_polaron} below). This correspondence emerges from the finite transferred photon momentum and thus is not affected by thermal shifts as observed in rf measurements \cite{Yan2019} (see Appendix~\ref{app:Ep_shift}). This is evident from the theoretical analysis in  Fig.~\ref{Fig:RamanSpectraTheory2}(a), where a comparison between the calculated polaron energy (solid blue) and the extracted peak position of the coherent Raman response (blue circles) is shown.
    \item \textbf{Background signal $P_{\mathrm{bg}}(\omega)$:} an asymmetric lineshape extending to high frequencies that contains the combined response arising from the incoherent part of the polaron, as well as from molecules that are dissociated by the Raman lasers. In a wavefunction picture, the former corresponds to contributions as given by the second term in Eq.\,\eqref{Eq:WaveFunctionPolaron}. We find that up to a rescaling by a factor, the shapes of the background spectra from polarons and molecules are similar.  
\end{enumerate}

Based on this identification, we are able to develop a fit model for the transition probability that is largely model-independent, and reflects the lineshape of the coherent and background signals
\begin{equation}\label{Eq_trans_prob_pairs_and_QP}
P(\omega)=\bar{Z} P_{\mathrm{coh}}(\omega;T_{p},\Ep^0,m^\ast)+(1-\bar Z)P_{\mathrm{bg}}(\omega;T_{\mathrm{bg}},E_b)\;.
\end{equation}
Here, both contributions $P_\text{coh}$ and $P_\text{bg}$ are normalized to unity. The `many-body polaron weight' $\bar Z$ quantifies the weight of the coherent polaron peak for a system with a finite density of impurities. In the limit  $T,x\to 0$, it reduces to the polaron quasiparticle residue $Z$ for interactions where the polaron is the ground state. Next, we determine suitable shape functions for $P_\text{coh}$ and $P_\text{bg}$.

\textbf{Raman coherent polaron peak.---} 
The Raman spectrum of a single polaron can be expressed as 
\begin{equation}\label{Eq_spectral_function_of_polaronsA}
\mathcal{A}_\mathrm{pol}(\omega,\veck)= Z \mathcal{A}_\text{coh}(\omega,\veck)+(1-Z)\mathcal{A}_\text{inc}(\omega,\veck)\;,
\end{equation}
where the coherent part can be approximated at low momenta as
\begin{equation}\label{Eq_spectral_function_of_polarons}
\mathcal{A}_\text{coh}(\omega,\veck)= \delta\left[\omega- \epsilon_{\kv + \dpv}+\Ep(\veck)\right]\;, 
\end{equation}
$Z$ is approximated as a momentum independent quasiparticle weight, and the dispersion $\Ep(\veck)=\Ep^0+\left.\veck^2\right/{2}m^\ast$ is parametrized by an effective mass, $m^\ast$, and the polaron energy, $\epsilon_\mathrm{pol}^0$. The function $P_\text{coh}$ accounts only for the Raman response arising from  $\mathcal{A}_\text{coh}(\omega)$ with $m^\ast$, $\epsilon_\mathrm{pol}^0$, and a polaron temperature $T_p$ being fit parameters. As  described below, the incoherent response arising from $\mathcal{A}_\text{inc}$ will be attributed to $P_\text{bg}$. 

Within this model it is useful to recognize that the total number of impurities can be interpreted as a sum of impurities contributing to the coherent response $N_\text{coh}$, as well as impurities in the incoherent polaron part $N_\text{inc}$ and molecules $N_{\mathrm{mol}}$,
\begin{equation}\label{Eq:NumberNI} 
    N_I = N_\text{coh} + (N_\text{inc}+N_\mathrm{mol})\;.
\end{equation}
The left-hand side, $N_I$, is responsible for the full signal, $P(\omega)$, while $N_\text{coh}= \bar{Z}N_I$ yields the contribution $\bar Z P_\text{coh}$ in Eq.\,\eqref{Eq_trans_prob_pairs_and_QP}. The sum $(N_\text{inc}+N_\mathrm{mol})=(1- \bar{Z}) N_I$, in turn, gives $(1-\bar Z)P_\text{bg}$.
Moreover, the number of polarons is given by $N_\mathrm{pol}=N_\text{coh}+N_\text{inc}$ with $N_\text{coh}=Z N_\mathrm{pol}$. Thus, within a model with momentum independent quasiparticle weight $Z\equiv Z_{\veck=0}$, one has $\bar{Z}=Z N_{\mathrm{pol}}/N_I$.
 
The coherent part of the polaron Raman spectrum $\mathcal{A}_\text{coh}$ is related to the coherent contribution of the full, many-impurity Raman response $\bar{\Gamma}(\omega) = \bar{\Gamma}_\text{coh}(\omega)+\bar{\Gamma}_\text{bg}(\omega)$ by 
\cite{PhysRevLett.98.240402,PhysRevA.80.023627}
\begin{equation}\label{Eq_total_raman_rate_coh}
\bar{\Gamma}_{\text{coh}}\left(\omega\right)=2\pi\Omega^2_e  \sum_\veck  \mathcal{A}_{\text{coh}}\left(\omega,\kv\right)\cdot n_F\left[\epsilon_{\mathrm{pol}}(\kv),T_p\right]\;,
\end{equation}
such that $\intop\mathrm{d}\omega\bar{\Gamma}_{\text{coh}} (\omega)= 2 \pi \Omega^2_e N_{\text{coh}}$.

Since the concentration of impurities is finite, polarons can be found at non-zero momenta \cite{PhysRevA.78.033614,Schmidt2011}. As Eq.\,\eqref{Eq_spectral_function_of_polarons} shows, a polaron with a momentum $\veck$ gives a coherent contribution to the Raman signal if $\Ep(\veck)=\varepsilon_{\veck+\qr}-\omega$, which can be solved for $\kq\left(\omega\right)\equiv\veck\cdot\hat{\qr}$. In particular, if $m^\ast=m$, this yields a linear relation between $\omega$ and $\kq$. Otherwise, the solution has a weak dependence on $\left(1-m/m^\ast\right)$ and   $k_{\perp}^2\equiv\veck^2-k_{\bar q}^2$. However, this dependence is only noticeable for $\kq$ close to $k_F$, and when the effective mass is substantially larger than the bare mass (see Appendix~\ref{app:fitting_val}). Neglecting this small effect, we obtain
\begin{equation}\label{Eq_defining_kq_for_polarons}
k_{\bar{q}}\left(\omega\right)=\frac{m}{\bar{q}}\left(\omega+\Ep^0\right)-\frac{\bar{q}}{2}\;, 
\end{equation}
where $\bar q\equiv|\qr|$.

Evaluation of Eq.\,\eqref{Eq_total_raman_rate_coh}  shows that the coherent polaron Raman rate is proportional to the one-dimensional momentum distribution of polarons in the direction of $\qr$, $\bar{\Gamma}_{\text{coh}}(\omega)=2\pi m\left. \Omega_e^2N_\text{coh} n_{\text{P}}\left[k_{\bq}\left(\omega\right)\right]\right/ \bar{q}$.
In the local density approximation (LDA) this distribution is given by (for details see Appendix~\ref{sect:appendix_raman_coherent})
\begin{equation}\label{Eq_momentum_distribution_of_minority}
n_\text{P}\left[k_{\bq}\left(\omega\right)\right]=-\frac{6T_p^{\frac{5}{2}}\left(\varepsilon_F-\Ep^0\right)^{-\frac{3}{2}}}{\sqrt{\pi}x\bar{Z}k_F\varepsilon_F m/m^{\ast}}\mathrm{Li}_{\frac{5}{2}}\left(-\zeta_{\text{P}}e^{-\frac{k_{\bar{q}}^2}{2m^{\ast}T_p}}\right)\;. 
\end{equation}
Here $x$ is the global impurity concentration, $\mathrm{Li}_{5/2}$ is the polylogarithm function, and $\zeta_{\text{P}}=e^{- (\Ep^0-\mu)/ T_p}$ is the fugacity of polarons. The chemical potential $\mu$ is tuned so that Eq.\,\eqref{Eq_total_raman_rate_coh} is normalized to the number of polarons that contribute to the coherent part of the response.

As a final step, in order to obtain the probability $P_\text{coh}$, the response $\bar{\Gamma}_\text{coh}$ has to be normalized to unity. We arrive at
\begin{equation}\label{EQ_trans_prob_QP}
P_{\mathrm{coh}}(\omega;T_{p},\Ep^0,m^\ast)=\frac{m}{\bar{q}}n_\text{P}\left[k_{\bq}\left(\omega\right)\right]\;.
\end{equation}

\textbf{Background Raman signal.---} 
We find that the experimental spectral lineshape arising from the incoherent polaron part and the molecules are fit
well by the response of a thermal gas of molecules (see Appendix~\ref{app:fitting_val}). Indeed, this model covers well the overall spectral weight of the background and allows us to incorporate the $\sim\omega^{-3/2}$ tail.

The fit function $P_\text{bg}(\omega)$ is derived by considering a pair of atoms bound as a molecule with a binding energy $E_b$ and a center-of-mass momentum $\mathbf{k}_{\mathrm{cm}}$. The Raman process dissociates the pair and changes the center-of-mass momentum to $\mathbf{k}_{\mathrm{cm}}+\qr$. In addition, the unbound fermions acquire a relative momentum $\krel\equiv|\veck_\text{rel}|$. Energy conservation yields
\begin{equation}\label{Eq_k_rel_squared}
\krel^2=m\omega-mE_b-\frac{\bar{q}^2}{4}-\frac{\bar{q}k_{\mathrm{cm},\bar{q}}}{2}\;,
\end{equation}
where $k_{\mathrm{cm},\bar{q}}\equiv\mathbf{\kcm}\cdot \hat{\dpv}$.

The probability $F\left(\mathbf{k}_{\mathrm{rel}}\right)$ that a pair will be dissociated with a relative momentum $\veck_\text{rel}$ is determined by its relative envelope wavefunction, resulting in \cite{PhysRevA.71.012713}
\begin{equation}\label{Eq_k_rel_distribution}
F\left(\mathbf{k}_{\mathrm{rel}}\right)=\frac{\pi^{-2}\sqrt{mE_b}}{\left(mE_b+k_{\mathrm{rel}}^2+k_{\mathrm{rel},\bar{q}}\bar{q}+\frac{\bar{q}^2}{4}\right)^2}\;.
\end{equation}
For $\mathbf{k}_{\mathrm{cm}}$ we assume a thermal Boltzmann distribution at temperature $T_{\mathrm{bg}}$, $G(\mathbf{k}_{\mathrm{cm}}, T_{\mathrm{bg}})=(4 \pi m T_{\mathrm{bg}})^{-3/2} e^{-\mathbf{k}_{\mathrm{cm}}^2 /4m T_{\mathrm{bg}}}$, which allows us to derive an analytical fit model. The combined probability to find a pair with an initial $k_{\mathrm{cm},\bar{q}}$, and final $k_{\mathrm{rel}}$, is then given by the product of $F$ and $G$. To obtain the Raman transition probability as a function of frequency, we change variables from $(k_{\mathrm{cm},\bar{q}},\, k_{\mathrm{rel}})$ to $(k_{\mathrm{cm},\bar{q}},\, \omega)$ using (\ref{Eq_k_rel_squared}). Finally,  integration over the angle $\hat{\mathbf{k}}_{\mathrm{rel}}\cdot\hat{\qr}$ and $k_{\mathrm{cm},\bar{q}}$ yields the normalized Raman transition probability for the background signal (for details see Appendix~\ref{mol_spectrum}),
\begin{eqnarray}\label{Eq_P_cpairs_full}
 & & P_{\mathrm{bg}}\left(\omega;T_{\mathrm{bg}},E_b\right)=\\
& & \sqrt{\frac{2E_{b}}{\pi^{3}T_{\mathrm{bg}}}}\intop_{-\infty}^{\left.2m\tilde{\omega}\right/q}\mathrm{d}k_{\mathrm{cm},\bar{q}}\frac{2m\sqrt{ 2m\tilde{\omega}-k_{\mathrm{cm},\bar{q}}\bar{q}}e^{-\frac{k_{\mathrm{cm},\bar{q}}^{2}}{4mT_{\mathrm{bg}}}}}{4mE_{b}\bar{q}^{2}+\left(\bar{q}^{2}+k_{\mathrm{cm},\bar{q}}\bar{q}-2m\omega\right)^{2}}\;,\nonumber
\end{eqnarray}
where $\tilde{\omega}\equiv\omega-E_b-\frac{\bar{q}^2}{4m}$. This integral does not have an analytic solution, but it can be readily calculated numerically.

Note that in this fit model, $T_{\mathrm{bg}}$ and $E_b$ are effective temperatures and binding energies. Since $P_\text{bg}$ also describes the incoherent polaron contribution, $E_b$ can be interpreted as the molecular binding energy $-\Em$ only in the limit of large $\kfainv$. The effective temperature $T_{\mathrm{bg}}$ compensates for the absence of Pauli blocking in the molecular model, and therefore should not be interpreted as the physical temperature of molecules.

\section{Experimental results}
\label{Sec:ExperimentalResults}

\begin{figure}
	\centering
	\includegraphics{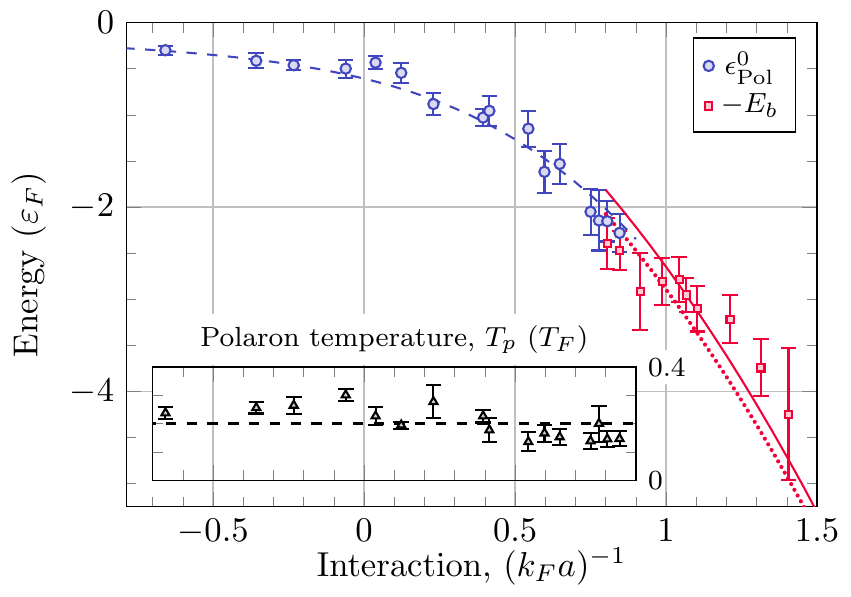}
	\caption{\textbf{Measured polaron and pairs binding energies.}  
	The polaron energies, $\Ep^0$, (blue circles) are obtained using Eq.\,\eqref{Eq_omega_0_of_polaron} from the position of the spectral peak, $\omega_0$. The theoretical prediction obtained from the variational Ansatz Eq.\,\eqref{Eq:WaveFunctionPolaron} is shown as a dashed blue line.
	The $E_b$ parameter (red squares) is determined by fitting the Raman spectra with Eq.\,\eqref{Eq_trans_prob_pairs_and_QP}. For $\kfainv>0.5$, it is in good agreement with the energy obtained from the simple molecular Ansatz Eq.\,\eqref{Eq:WaveFunctionMolecule} (solid red line). The dotted red line shows the result of an improved molecular Ansatz \cite{Punk2009}.
	Note that the theoretical polaron and molecule energies are averaged over the harmonic trap (see Appendix~\ref{appendix_local_density_approximation}), and as a result, they cross at a $\kfainv$ slightly lower than the predicted polaron-to-molecule transition in a homogeneous gas. Inset: extracted polaron temperatures. The approximate majority temperature is denoted by the dashed line.
	}
	\label{fig:polaron_and_molecule_energies}
\end{figure}

The peak of the coherent polaron spectrum, as given in Eq.\,\eqref{EQ_trans_prob_QP}, is at $\kq=0$. According to Eq.\,\eqref{Eq_defining_kq_for_polarons}, this maximum is attained for 
\begin{equation}\label{Eq_omega_0_of_polaron}
\omega_0=\frac{\bar{q}^2}{2m}-\Ep^0\;. 
\end{equation}
For interactions below the transition point, the most significant contribution to the spectral peak stems from the coherent part of  polarons.
Thus, from $\omega_0$ we can determine the polaron energy, $\Ep^0$. We find the peak position (dotted vertical lines in Fig.~\ref{fig:raman_data_for_polarons}) by fitting the points above the median with a skewed Gaussian \cite{Stancik2008}. The resulting polaron energies are plotted in Fig.~\ref{fig:polaron_and_molecule_energies} (blue circles). We compare the data and find excellent agreement with the predictions of our theoretical model (dashed line), which in turn are close to diagMC and T-matrix calculations \cite{Combescot2007,Prokofev2008,Prokofev2008a,Punk2009,Chevy2010}. Beyond $\kfainv=0.9$ the weight of the coherent peak is small. Thus we restrain the fit in this regime by using the  value for $\Ep^0$ obtained from the Chevy Ansatz.

Next, we extract the polaron  weight $\bar{Z}$ by fitting the measured spectra with Eq.\,\eqref{Eq_trans_prob_pairs_and_QP}.
The effective temperature parameter, $T_{\mathrm{bg}}$, controls the sharpness of the background spectrum onset.
We fix $T_{\mathrm{bg}}=2\,T_F$, which yields a minimal systematic error in extracting $\bar{Z}$ (see Appendix~\ref{app:fitting_val}). The polaron effective mass, $m^\ast$, is strongly coupled to the polaron temperature. To make the fit robust, we set $m^\ast$ to the trap-averaged theoretical value at $k\rightarrow0$, calculated from Eq.\,\eqref{Eq:WaveFunctionPolaron}. We find that the effective mass  modifies the extracted polaron weight and molecular binding energy only marginally. In fact, setting $m^{\ast}$ to the bare mass leads to a maximal deviation of less than $0.4\,\sigma$. We are left with three free fitting parameters: $\bar{Z}$, $E_b$, and $T_p$.

Examples of fits are shown in Fig.~\ref{fig:raman_data_for_polarons} (solid lines). Overall, we find excellent agreement between the fits and the measured spectra throughout the whole interaction range.
The light and dark shaded areas beneath the curves are the spectral contributions of the coherent part of the polaron and the background, respectively.
The BCS-side data (blue circles) are dominated by a nearly symmetric quasiparticle peak with $\bar{Z}=0.91(3)$, while the BEC-side data (black triangles) are dominated by the asymmetric pair dissociation spectra leading to a small coherent weight $\bar{Z}=0.18(2)$. The unitary data (red squares) shows both the symmetric peak and an asymmetric tail. The quasiparticle weight is $\bar{Z}=0.58(3)$, close to the value of $0.47(5)$ which was measured for $^{6}$Li atoms with rf spectroscopy \cite{PhysRevLett.102.230402}.

\begin{figure}
	\centering
	\includegraphics{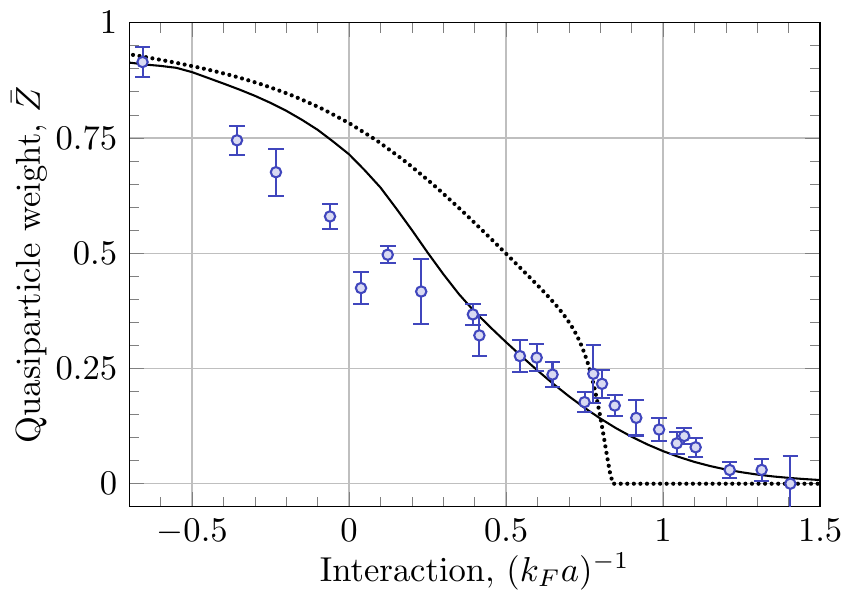}
	\caption{\textbf{Measured quasiparticle weight.} The quasiparticle  weight, $\bar{Z}$, is shown for different interaction strengths. Blue circles mark the weight extracted from the Raman spectra by identifying the nearly symmetric spectral lineshape arising from the coherent polaron contribution. $\bar{Z}$ is smoothly decreasing towards the  polaron--molecule transition, in agreement with our theoretical prediction, averaged over the harmonic trap using the LDA (solid line). For comparison, we also present the trap-averaged prediction for a single impurity at $T=0$ (dotted line) obtained from Ref.~\cite{Punk2009}.}
	\label{fig:quasiparticle_residue}	
\end{figure}

The coherent polaron spectral weight, $\bar{Z}$, extracted from the fits is shown in Fig.~\ref{fig:quasiparticle_residue}. It approaches unity on the far BCS side, as expected. For increasing $\kfainv$, we observe a smooth decrease of $\bar{Z}$. We compare the result to the calculation of our theoretical model in the LDA (solid line) and find good agreement for $\kfainv>0.4$. Indeed, as shown in Section~\ref{sect:Theory2}, it is essential to account for the coexistence of polarons and molecules. To demonstrate the crucial role of the finite impurity density and temperature, we also plot in  Fig.~\ref{fig:quasiparticle_residue} the prediction for a single impurity at zero temperature in the LDA (dotted line).
Our data clearly disagrees with this result, and in particular, does not exhibit a sudden change at the polaron-to-molecule transition as predicted in the single polaron limit.

The polaron temperature parameter, $T_p$, is shown in the inset of Fig.~\ref{fig:polaron_and_molecule_energies}. We find $T_p$ to be around $0.25\,T_F$, slightly higher than the measured majority temperature of approximately $0.2\,T_F$ (marked by a dashed line). A moderate systematic decrease of the extracted temperature is visible as the interaction strength is increased. We attribute this behavior to the reduction in the quasiparticle population due to a lower quasiparticle lifetime at high momenta.

We now turn to examine the $E_b$ parameter, which is extracted from the background signal of the Raman spectra. The results are presented as red squares in Fig.~\ref{fig:polaron_and_molecule_energies}. We also show the theoretical predictions obtained from the simple variational Ansatz (Eq.\,\eqref{Eq:WaveFunctionMolecule}) in the LDA (solid red). Including particle-hole dressing of the molecular state \cite{Punk2009} leads to a further lowering of the molecule energy (dotted red line). Since in the far BEC limit, the molecules dominate the Raman response, the parameter $E_b$ regains its physical interpretation as the molecular binding energy (up to a contribution on the order of $\epsilon_F$ arising from the neglect of the presence of a Fermi surface). In this region, we find good agreement between the data and the theoretical predictions.

\begin{figure}
	\centering
	\includegraphics{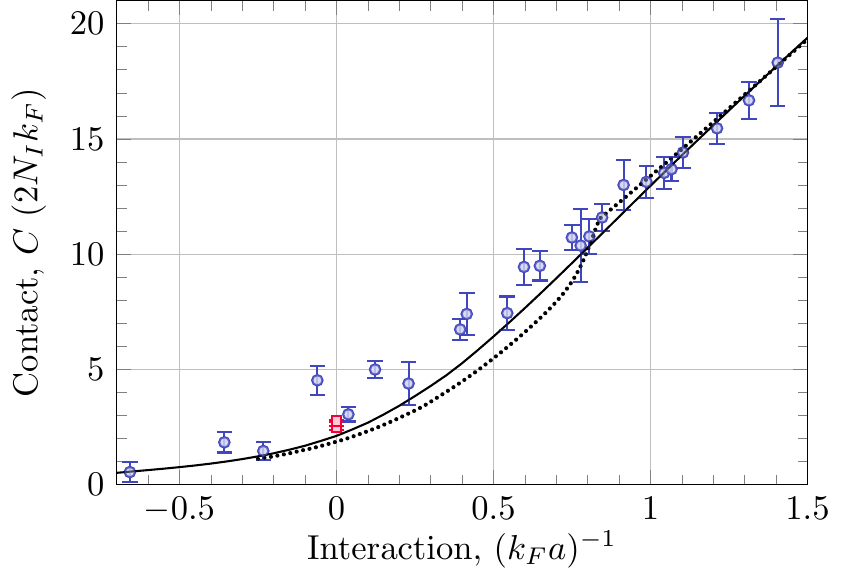}
	\caption{\textbf{Measured Tan contact.} The contact  coefficient ${C}$ is shown for different interaction strengths, obtained using Eq.\,\eqref{Eq:contact_from_Eb_and_Z} with $E_b$ and $\bar{Z}$ extracted from the Raman spectra.
	The theory for a single impurity at zero temperature predicts a discontinuous change in $C$ at the polaron-to-molecule  transition \cite{Punk2009}. While this discontinuity is smoothed when averaged for a harmonically trapped gas using the LDA (dotted line), an abrupt change is expected to remain. The data  significantly deviates from this prediction. Instead, it agrees well with our calculation taking into account the finite temperature and coexistence of polarons and molecules (solid line), see Section \ref{sect:Theory2}.	The two red squares indicate data  measured by the MIT group \cite{Yan2019} using rf spectroscopy of a  unitary, homogeneous $^6$Li gas at $T=0.17\,T_F$ (lower point) and $T=0.29\,T_F$ (upper point).
	}
	\label{fig:contact}	
\end{figure}

Finally, we extract the Tan contact, $C$, from our experimental data. The contact coefficient is related to the tail of the momentum distribution of the quasiparticles and measures the short-distance correlations between bath and impurity particles \cite{Tan08,Tan08a,Tan08b,Braaten08,PhysRevA.78.053606,Werner2012,Werner2012b,Rossi2018}. Moreover, the contact relates the high-momentum tail to various many-body quantities, such as the thermodynamic pressure and  quantifies the spectral weight in the universal $\omega^{-3/2}$ tail of the Raman spectra \cite{Zwerger_book,Nishida}.  

Being related to the derivative of the ground state energy, in the single-impurity limit at $T=0$, $C$ is expected to jump at the polaron-to-molecule transition \cite{Punk2009}. In our fitting model, the tail appears in the spectral contribution of incoherent polarons and molecules. Thus the contact is related to the parameters of our model by \cite{PhysRevA.78.053606,PhysRevLett.114.075301}
\begin{equation}\label{Eq:contact_from_Eb_and_Z}
C=4\pi(1-\bar {Z})\sqrt{\left.E_b\right/2\EF}\;.
\end{equation}
Here $C$ is given in units of $2 N_I k_F$ such that the high-frequency tail of the spectrum approaches $P(\omega)\to C\sqrt{\EF^{\phantom{1}}}\left/\sqrt{2}\pi^{2}\omega^{3/2}\right.$ \cite{PhysRevA.81.021601}.

The results for $C$ are shown as blue circles in Fig.~\ref{fig:contact}. For comparison, we plot the trap-averaged theoretical prediction for the contact in both the single-impurity limit (dotted line) and in the many-impurity case (solid line), as discussed in the following section. The data is in excellent agreement with the many-body model, and, in particular, it does not show any sudden change as predicted in the single impurity limit. We also indicate the contact measured by the MIT group with a homogeneous $^6$Li gas using rf spectroscopy \cite{Yan2019} (red squares), which agrees with our measurements to within the experimental uncertainty.

\section{Theoretical Raman spectra}\label{sect:Theory2}
As shown in the previous section, we find no experimental evidence for a discontinuity in the extracted observables.
We now demonstrate that this observation is consistent with a finite impurity density theory, which inherently features a first-order transition in the single-impurity limit. In fact,  discontinuities predicted in this limit are smoothed out by a finite impurity density. This effect becomes further amplified at finite temperature.  

In order to incorporate the finite impurity density and temperature in the calculation of Raman spectra, we adopt an effective quasiparticle approach. In this model, quasiparticle states ---obtained in the single-impurity limit--- are occupied thermally according to their quantum statistics. More precisely, we consider the polaron and molecule states, given by Eqs.\,\eqref{Eq:WaveFunctionPolaron} and~\eqref{Eq:WaveFunctionMolecule}, to be populated according to the Fermi-Dirac and Bose-Einstein distributions $n_{F/B}(\epsilon,T)=\left(\exp\left[\left.\left(\epsilon-\mu\right)\right/T\right]\pm 1 \right)^{-1}$, respectively. 
Here,  $\mu$ denotes the chemical potential which determines the impurity density at temperature $T$ via
\begin{equation}
n_{I}(\mu, T)= \frac{1}{V} \sum_\pv  \left(n_F\left[\Ep(\pv),T\right]+ n_B\left[\Em(\pv),T\right]\right)\;.
\label{Eq:ImpDensity}
\end{equation}
Note that the impurity temperature and chemical potential $\mu$ are set independently of the bath. 
Specifically, for all calculations in this section, $\mu$ is tuned to yield an impurity density of $n_I=0.15\,n$ at a finite impurity temperature $T=0.2\,T_F$ \footnote{The value of $n_I=0.15\,n$ is taken as the typical value of the experimentally realized $\braket{x}$.}.

\begin{figure}
	\centering
	\includegraphics{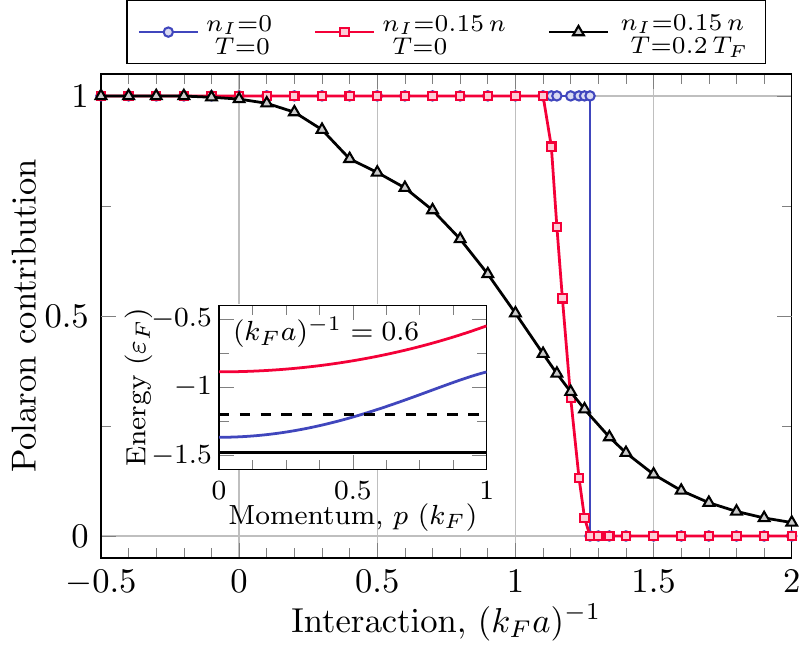}
	\caption{\textbf{Calculated polaron contribution.} Fraction of impurity particles propagating as polarons as a function of interaction strength $\kfainv$. The impurity particles which are not polaronic are bound to a bath particle, thus forming a molecule. In the single-impurity limit and $T=0$ (blue), there is a sharp transition between a polaron and a molecule at $\kfainv_c\approx 1.27$. Finite impurity density ($0.15\, n$, red) leads to smoothing of the transition for $\kfainv<\kfainv_c$. When the temperature is increased ($0.2\, T_F $, black), the polaronic branch is populated also for $\kfainv>\kfainv_c$ and the sharp transition disappears. The inset shows the polaron (blue) and molecule (red) dispersions for $n_I=0.15\,n$ and $\kfainv=0.6$, along with the chemical potential at $T=0$ (dashed black) and $T=0.2\,T_F$ (solid black). At $T=0.2 \, T_F$, polarons and molecules are populated only by thermal excitations, while at $T=0$ the chemical potential is above the minimum of the polaron dispersion such that a well-defined polaronic Fermi surface forms.
     }
    
     \label{Fig:ContributionsTheory}
\end{figure}

In Fig.~\ref{Fig:ContributionsTheory}, the polaron contribution to the total impurity density in the initial state is shown as a function of the interaction strength $\kfainv$. Evidently, in the single-impurity limit at zero temperature (blue circles) the system undergoes a sharp transition from a purely polaronic to a purely molecular state at $\kfainv\approx 1.27$. 

Still at zero-temperature but at finite impurity density (red squares), the system is purely polaronic up to $\kfainv\approx 1.1$. At this interaction strength the chemical potential reaches the minimum of the molecular dispersion and, henceforth, it remains pinned to that value (for an illustration of this effect, we show in the inset of Fig.~\ref{Fig:ContributionsTheory} the dispersion relations of polarons (blue) and molecules (red) as well as the impurity chemical potential $\mu$ (black); for a more detailed discussion see Appendix \ref{sect_app_chempots}). Accordingly, for $1.1 \lesssim \kfainv\lesssim1.27$ molecules begin to condense in the lowest-lying molecular state while the polaron Fermi surface shrinks and eventually vanishes at the polaron-to-molecule transition. In this range of $\kfainv$, polarons and molecules coexist, even at $T=0$. Beyond the transition, the system forms a molecular condensate within the bath of the remaining majority atoms. 

Finally, at finite temperature (black triangles) polarons and molecules coexist as a thermal mixture. This blurs the transition and leads to a smooth interpolation between polaron and molecule dominated regimes. Note that the temperatures considered in this work exceed the critical temperature for Bose-Einstein condensation of molecules, which thus form a purely thermal gas.

In our calculations, we only occupy quasiparticle states with an infinite lifetime and finite quasiparticle weight. This ensures that the initial state has an infinite lifetime as expected for an equilibrium state, and also that the quasiparticle picture remains valid. As a consequence, we cut off the quasiparticle populations of the polaronic and molecular states at momenta where they no longer feature poles on the real frequency axis. 

For the polaron, this momentum cutoff occurs when the energy becomes positive or $|\alpha^{\pv}_0|^2$ vanishes. For the molecule, however, this cutoff occurs when the dispersion intersects the continuum of states delimited by a parabola of the form $\left(|\vecp|- k_F\right)^2$. In fact, this condition causes the slight dent in the polaron contribution  at $\kfainv\approx 0.4$ visible in Fig.~\ref{Fig:ContributionsTheory}, as beyond that value  molecules have a well-defined dispersion for all momenta and thus do not have a cutoff. Similarly, for the polaron its cutoff condition changes at around $\kfainv\gtrsim 0.3$. 
We note that in order to accurately incorporate states with a finite lifetime or continuum states, a solution of the full imbalanced problem would be necessary.

As discussed in Section~\ref{sect:FermiPolaronModel}, we compute the Raman spectra given by Eq.\,\eqref{Eq:FullRamanSpectrum} for finite impurity density and temperature by summing the single-impurity Raman spectra over all impurity momenta, weighted by their occupation probability. 
The single-impurity Raman spectra are obtained by computing the matrix elements $\braket{ f|\hat V_R|i}$ in Eq.\,\eqref{Eq:FermisGoldenRule} for the Ans{\"a}tze Eqs.\,\eqref{Eq:WaveFunctionPolaron} and \eqref{Eq:WaveFunctionMolecule}. This yields
\begin{align}\label{Eq:RamanPolaronCohAndInc}
\mathcal{A}(\omega,\psi_P^\pv)= &\big|\alpha_{0}^\pv \big|^2 \delta\left[\omega-\varepsilon_{\pv +\dpv} + \Ep(\pv)\right] \\
+\sum_{\kv,\qv }{}^{'} &\big|\alpha_{\kv,\qv}^\pv \big|^2 \delta\left[\omega-\varepsilon_{\pv +\dpv+ \qv -\kv}-  \varepsilon_{\kv}+ \varepsilon_{\qv} + \Ep(\pv)\right]  \nonumber
\end{align}
for the polaron, and
\begin{align}\label{Eq:RamanMolecule}
&\mathcal{A}(\omega ,\psi_M^\pv)=   \\
&\qquad\sum_\kv{}^{'}\big|\beta_\kv^\pv\big|^2\delta[\omega-\varepsilon_\kv-\varepsilon_{\pv-\kv+\dpv}+\Em(\pv)+\varepsilon_F]
\nonumber
\end{align}
for the molecule. 
These expressions make explicit the three contributions that make up the many-body Raman spectrum as we have discussed in the previous sections, namely a coherent and incoherent polaron part as well as a molecular part. Note that our Raman spectra 
are normalized such that they sum to unity once integrated over frequency $\omega$. As discussed in Section~\ref{sect:FermiPolaronModel}, in Fig.~\ref{Fig:RamanSpectraTheory2}, we exemplarily show many-body Raman spectra for three $\kfainv$ across the transition. In the following, we describe how such Raman spectra give access to  quasiparticle properties in the regime of finite impurity concentration.

The polaron $Z$-factor can be obtained from the self-energy of the impurity via $Z_\pv=\left|1-\partial_\omega \Sigma(\omega,\pv)|_{\omega=\omega_\pv}\right|^{-1}$, where $\omega_\pv$ is a pole in the retarded Green's function of the quasiparticle at momentum $\pv$ \cite{Abrikosov1975}. The momentum-dependent weight $Z_{\pv}$ can, alternatively, be obtained from the overlap of the non-interacting wavefunction with the interacting one, $Z_\pv=|\alpha_0^{\pv}|^2$ \footnote{If the polaron is a stable quasiparticle ($\text{Im}(\omega_\pv)=0$), its $Z$-factor within the polaron Ansatz (\ref{Eq:WaveFunctionPolaron}) is given by $|\alpha_0^{\pv}|^2$. These definitions are, however, not equivalent if the polaron acquires a finite lifetime ($\text{Im}( \omega_\pv)\neq0$). This discrepancy can be seen from comparing the definition of $Z_\pv$ using the self-energy to the normalization condition for $|\alpha_0^{\pv}|^2$ as they differ in the placement of the absolute value bars.}. In the molecular state, the impurity is bound to a bath particle, leading to a vanishing $Z$-factor in the thermodynamic limit \cite{Punk2009}.

Similar to the single-impurity quasiparticle residue, $\bar{Z}$ is given by the spectral weight of the coherent part of the Raman spectra (blue-shaded area in Fig.~\ref{Fig:RamanSpectraTheory2}). It can be calculated from the single impurity residues via
\begin{equation}
    \bar{Z}=\frac{1}{N_I} \sum_\pv Z_\pv \cdot n_F\left[\Ep(\pv),T \right] \; .
    \label{Eq:ManyBodyZ}
\end{equation}

\begin{figure}
	\centering
	\includegraphics{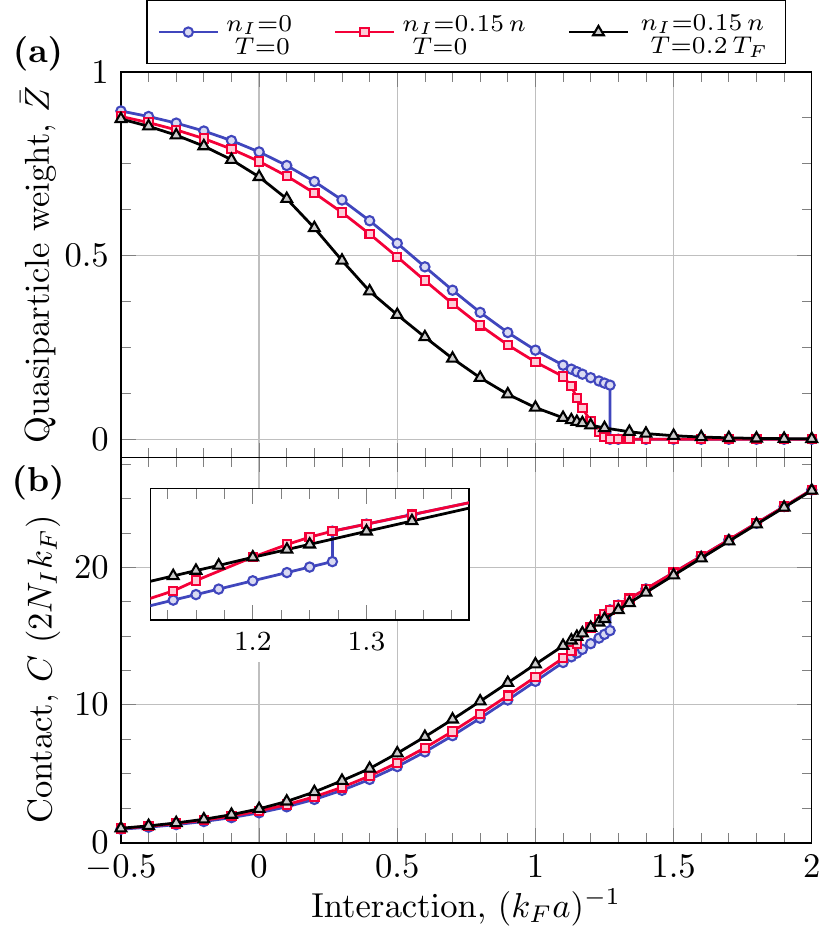}
	\caption{\textbf{Calculated quasiparticle weight and contact.} The quasiparticle weight \textbf{(a)} and contact coefficient \textbf{(b)} are shown for different interaction strengths. In the single-impurity limit (blue), the transition between polaron and molecule at $\kfainv\approx 1.27$ leads to a sharp jump between the polaronic and molecular residues and contacts. As in Fig.~\ref{Fig:ContributionsTheory}, at $T=0$ and finite impurity density ($0.15\, n$, red) the transition is smoothed, and eventually blurred at finite impurity density  and temperature ($0.2\, T_F $, black). The inset in the lower figure shows a magnification around the transition point.
	}
\label{Fig:ResidueAndContact}
\end{figure}

As evident in Fig.~\ref{Fig:ResidueAndContact}(a), in the single-impurity limit $\bar{Z}$ features a sharp jump at the polaron-to-molecule transition where it drops to zero as the polaron is not populated anymore. Importantly, in this limit $\bar Z$  reduces to the zero-momentum polaron residue, $\bar Z = Z_\textbf{0}$, before the transition.
At finite impurity density and $T=0$, this jump is smoothed with the many-body weight again dropping  to zero at the transition. 

At finite temperature and density, the transition is completely blurred with $\bar{Z}$ being lowered on the polaronic side compared to $T=0$ and the single impurity limit. This is due to the circumstance that, first, some impurity particles are propagating as molecules with a vanishing residue and, second, also finite-momentum polarons with a lower residue $Z_{\pv}< Z_{\pzero}$ contribute to the many-body weight $\bar Z$. 

As shown in Fig.~\ref{fig:quasiparticle_residue}, the predicted smooth behavior of the many-body weight $\bar Z$ is
consistent with the experimental observation.
The overestimation of the theoretical values for $\bar Z$ in the polaron-dominated interaction regime for $\kfainv\lesssim 0.4$ can be attributed to several reasons. Firstly, the single-impurity polaron weight $Z$  will be reduced when higher-order terms are included in the wavefunction Ansatz Eq.\,\eqref{Eq:WaveFunctionPolaron}. Secondly, due to the neglect of finite-lifetime molecular states, the polaron contribution in the initial state is overestimated.
Thirdly, the disregard of finite-lifetime polarons leads to an effective  population transfer to low-momenta polaron states which, again, results in a higher quasiparticle weight. 

The large-frequency behavior of the single-impurity and many-body Raman spectra is governed by a power-law proportional to the Tan contact $C$ (see Section \ref{Sec:ExperimentalResults}). Based on the single-impurity coefficients (Appendix~\ref{Appendix:SingleImpContact}), the many-impurity contact coefficient of the full many-body spectrum is determined by 
\begin{align}
     C=&\sum_\pv C_{\mathrm{pol}}[\pv,\Ep(\pv)]\cdot n_F\left[\Ep(\pv),T \right]\nonumber\\
    &+ \sum_\pv C_{\mathrm{mol}}[\pv,\Em(\pv)]\cdot n_B\left[\Em(\pv),T \right]\;.
    \label{Eq:ManyBodyContact}
\end{align}

The contact $C$ is shown in Fig.~\ref{Fig:ResidueAndContact}(b) as a function of $\kfainv$ at finite impurity density for $T>0$ and $T=0$, along with the prediction in the single-impurity limit at $T=0$. As can be seen, these scenarios differ significantly only around the polaron-to-molecule transition. While the single-impurity limit features a discontinuity, already the finite density graph at $T=0$  shows a smooth transition between the polaronic and molecular contacts. At finite temperature this transition is further blurred. This is in line with the experimental observation shown in Fig.~\ref{fig:quasiparticle_residue}, where the measured data are compared to the trap-averaged, theoretical prediction for $C$ (solid black line).

\section{Discussion}\label{Discussion}

In this work, we have investigated the attractive Fermi polaron problem at finite impurity density and temperature, employing a novel Raman spectroscopy technique. The main advantage of this approach compared to rf spectroscopy is that the momentum transfer imparted by the two-photon transition is significant relative to the atomic momentum. As a result, Raman spectroscopy allows us to \emph{directly} probe the previously inaccessible momentum distribution of polarons. In order to maintain a good signal to noise ratio when working at a low impurity density, we additionally employ a high-sensitivity fluorescence detection scheme with which we can reliably measure signals of only a few atoms \cite{Shkedrov2018}. This allows us to probe the polaron-to-molecule transition at finite impurity density in previously unattainable regimes.

To extract physical quantities from the data, we have developed a simple fitting model that leverages the separation of the Raman spectra into two contributions: the nearly symmetric coherent polaron response, and an asymmetric background arising from the incoherent response of polarons and from molecules. From the measured Raman spectra we obtain the polaron energy, the quasiparticle spectral weight, and the contact parameter. 

In order to gain a better understanding of our measurements, we have devised a theoretical model based on a variational description of polarons and molecules that takes into account finite impurity density and temperature. The physical picture that arises from our experimental and theoretical observations is intriguing: All measured quantities show a smooth transition with no sudden changes around the predicted polaron-to-molecule transition. As we show theoretically, this is explained by the population of polarons and molecules at finite momenta, resulting from the finite impurity density and temperature. The excellent agreement between the theoretical model and the experimental data strongly suggests a coexistence phase of polarons and molecules around the interaction strength where the first-order transition in the single-impurity limit takes place. We stress that this coexistence region and the smooth transition from polarons to molecules is a general characteristic of any realistic scenario where many impurities are present.

To interpret our experimental data we have employed a quasiparticle theory, in which the many-body Hilbert space of impurity particles is spanned by single-particle states obtained from  variational wavefunctions. The approximate nature of this approach is reflected by the fact that polarons and molecules are effectively created by composite impurity-bath operators that maintain their respective commutation relations only approximately. Correspondingly, both impurity-induced correlations between majority fermions, as well as quasiparticle interactions (polaron--polaron, polaron--molecule, and molecule--molecule), induced by the Fermi sea, are neglected. The accurate inclusion of such correlation effects, which could, for instance, describe an instability of a Fermi polaron gas towards $p$-wave superfluidity \cite{Bulgac20062,Nishida2009}, presents a formidable theoretical challenge. While the finite density of impurities can be included in quantum field theory approaches, the systematic study of polaron--polaron interactions requires the inclusion of extended sets of vertex functions \cite{Camacho2018,Rossi2018b,frank2018} that are beyond the reach of mean-field approximations. Similarly, the development of wavefunction-based approaches that systematically include a finite number of impurities is challenging. Here, a major task is the inclusion of higher-order particle-hole excitations that become crucial not only in order to account for induced correlations, but also to ensure that the polaron dressing of each individual impurity is fully accounted for \cite{Li2020}. 

While the development of such approaches remains an outstanding challenge, it holds promise to shed further light on the nature of the phase diagram of highly-imbalanced quantum gases \cite{PhysRevLett.100.030401,zwerger2016strongly,frank2018}.
Our findings suggest that close to the interaction where the polaron-to-molecule transition takes place, polarons and molecules coexist when the temperature is above the critical temperature of molecular Bose-Einstein condensation. At lower temperatures, the phase diagram is not yet understood and contrasting predictions have been made. On the one hand, at zero-temperature, the polaron-to-molecule transition marks the endpoint of a fermionic polaron phase, where its finite Fermi surface volume vanishes and a polarized superfluid phase consisting of molecules is expected to take over  \cite{Sachdev2006,Punk2009}. On the other hand, considering the strong atom-dimer interactions close to the transition point \cite{Petrov2005}, it has been predicted that the system might become unstable towards phase separation between superfluid and normal phases \cite{PhysRevLett.100.030401}. The application of Raman spectroscopy in an imbalanced Fermi gas at lower temperature \cite{Grimm2017} and homogeneous traps \cite{Mukherjee2017,Moritz2018,Mukherjee2019,Bohlen2020} might help to distinguish these scenarios and allow one to experimentally determine, e.g., the transition temperature towards  phase separation. 
 
Furthermore, away from the transition point a plethora of phases has been discussed in the literature, ranging from $p$-wave pairing of polarons to the FFLO phase \cite{PhysRev.135.A550,Larkin_Ovchinnikov_1964,Zwerger_book,zwerger2016strongly}. Interestingly,  our results in the impurity limit already hint at some of these possibilities. As shown in Fig.~\ref{Fig:RamanSpectraTheory2}, we find that excited molecules in the polaronic regime feature a dispersion relation with a minimum at finite momentum (see also Refs.~\cite{Schmidt2012b,schmidt2013thesis} and a recent discussion in Ref.~\cite{Cui2020}). This effect may be regarded as a precursor of the  long-sought-after FFLO phase  in the imbalanced BEC-BCS crossover \cite{Son2006,PhysRev.135.A550,Larkin_Ovchinnikov_1964,Hulet_nature_2010,Piazza2016}, which  emerges due to the macroscopic occupation of such molecular states at finite momentum. Intriguingly, our calculation of the momentum-resolved  spectral function of polarons shows that they as well feature a roton-like minimum at finite momentum in their excited state. This raises the question of whether these finite-momentum states can be prepared in a controlled way, which could subsequently lead to the formation of a metastable, non-equilibrium polaron gas with a non-trivial Fermi surface topology \cite{KangSachdev2006,Sachdev2018}.

One possible way to study these questions is the extension of Raman spectroscopy to Raman \emph{injection} spectroscopy. Similar to rf injection spectroscopy \cite{Regal2003,Cheuk2012,Kohstall2012,Koschorreck2012}, the system is initially prepared in a weakly-interacting state and driven to a state where impurities strongly interact with their environment. Thus, with Raman injection one can prepare polarons at a specific momentum.  This enables a direct measurement of key polaron properties, such as the momentum-dependent effective mass, residue and lifetime. In addition, Raman injection could potentially facilitate the population and observation of the elusive finite-momentum polaron and molecular states as precursors of exotic phases in the BEC-BCS crossover. Moreover, Raman injection may also provide a promising means to probe unoccupied excitation branches of the spectral function \cite{PhysRevA.80.023627}. Such experiments could enable the controlled study of the momentum relaxation rate of polarons, which is currently investigated as a pathway for the realization of polaron-polariton-induced optical gain in two-dimensional semiconductor heterostructures \cite{Tan2020}.

Finally, in cold atom experiments, both fermionic and bosonic impurities can be implemented. This opens the exciting perspective of  studying  the fate of the polaron-to-molecule transition in highly imbalanced Bose-Fermi mixtures. This question has recently become the focus of experimental and theoretical studies  of interacting exciton-electron gases in two-dimensional transition-metal dichalcogenides \cite{Mak2013}, where the coexistence between molecular exciton-electron bound trion states and Fermi-polarons may lead to novel electronic  and optical properties \cite{Sidler2016,Efimkin2017,Cotlet2019,Glazov2020,imamoglu2020,Fey2020}.

\begin{acknowledgments}
We thank Felix Werner and Wilhelm Zwerger for inspiring discussions and valuable input.
This research was supported by the Israel Science Foundation (ISF), grant No. 1779/19, and by the United States - Israel Binational Science Foundation (BSF), grant No. 2018264. R.~S. is supported by the Deutsche Forschungsgemeinschaft (DFG, German Research Foundation) under Germany's Excellence Strategy -- EXC-2111 -- 390814868. G.~N. is supported by the Helen Diller Quantum Center at the Technion. O.~K.~D. and J.~v.~M. are supported by a fellowship of the International Max Planck Research School for Quantum Science and Technology (IMPRS-QST).
\end{acknowledgments}

\FloatBarrier
%


\appendix

\section{Local density approximation for spin-imbalanced gas}\label{appendix_local_density_approximation}
To account for the non-uniform atomic density, we compare measurements to average quantities calculated in the LDA. To this end, we assume that the distribution of the majority atoms is not affected by the presence of the minority atoms; hence it can be calculated as for non-interacting fermions. The minority density distribution, $n_I\left(\mathbf{r}\right)$, is calculated by taking into account the interactions with the majority atoms through a renormalization of the confining potential: $V_2\left(\mathbf{r}\right)=V_1\left(\mathbf{r}\right)(1-\Ep^0/\EF)$, where $V_1\left(\mathbf{r}\right)$ is the potential felt by the majority atoms \cite{Lobo2006}. We neglect the weak interactions between polarons \cite{Mora2010}. The expected value of any observable, $A$, is then given by the minority-weighted local density average: $\langle A \rangle=\frac{\int\mathrm{d}^3\mathbf{r} A\left(\mathbf{r}\right)n_I\left(\mathbf{r}\right)}{\int\mathrm{d}^3\mathbf{r} n_I\left(\mathbf{r}\right)}$. Notice that when we compare experimental results to theory as a function of $\kfainv$, the Fermi wave vector $k_F$ is that of the trap, namely the local $k_F$ at the center of the trap.

\section{Dependence of the coherent spectrum peak on the transferred photon momentum}\label{app:Ep_shift}
In this appendix, we examine the correlation
between the peak position of the coherent polaronic spectrum and the energy of the zero-momentum attractive polaron. As experimentally observed in Ref.~\cite{Yan2019}, at finite impurity density there is a temperature-dependent shift of the rf spectrum peak position relative to the zero-momentum polaron energy. Our theoretical model captures correctly this phenomenon and shows that it is absent in Raman spectroscopy.

\begin{figure}
	\centering
	\includegraphics{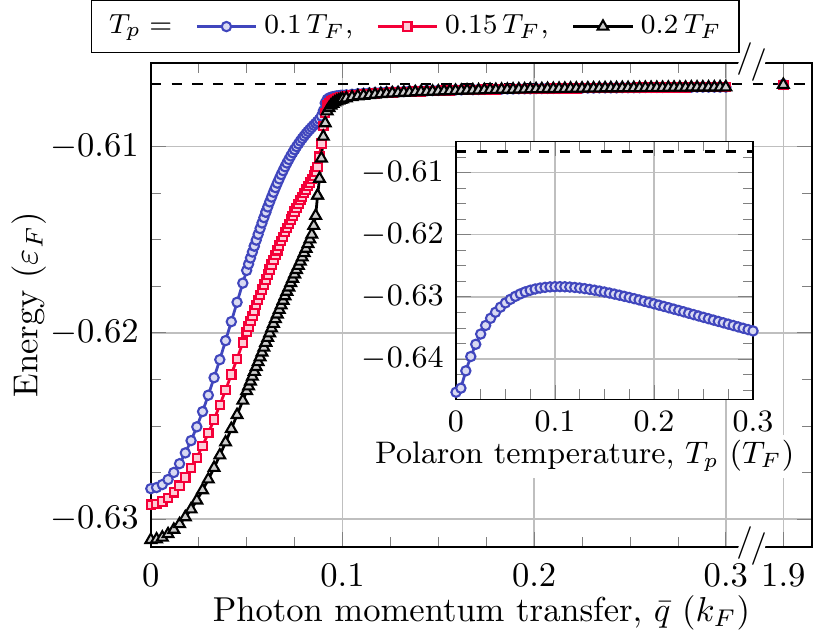}
	\caption{\textbf{Coherent spectral peak position at unitarity.} Using the full many-body model for the Raman spectra (first term of Eq.\,\eqref{Eq:RamanPolaronCohAndInc}), we plot the coherent peak position vs. transferred photon momentum for polaron temperatures $T_p=0.1\,T_F$ (blue circles), $0.15\,T_F$ (red squares), and $0.2\,T_F$ (black triangles) at $\kfainv=0$ with $n_I=0.15\,n$. We observe convergence to the zero-temperature value $\Ep=-0.6066\,\EF$ (dashed line) for photon momentum transfer larger than $0.1k_F$. Specifically, with Raman spectroscopy in our experiment ($\bq=1.9\,k_F$), no shift is expected. Inset: temperature dependence of the spectral peak position with conventional rf spectroscopy ($\bar{q}=0$).}
	\label{fig:peak_position_vs_q}
\end{figure}

In order to calculate this shift, we compute the peak position of the coherent polaron contribution (first term of Eq.\,\eqref{Eq:RamanPolaronCohAndInc} within Eq.\,\eqref{Eq:FullRamanSpectrum}) as a function of the photon transfer momentum. The results at unitarity, shifted by the recoil energy, are shown in Fig.~\ref{fig:peak_position_vs_q} for three relevant temperatures. In rf spectroscopy ($\bq\rightarrow 0$), we find that the peak is shifted to energies lower than the zero-momentum polaron energy ($\Ep$, dashed line). As the photon transfer momentum increases, however, this shift rapidly vanishes. Already for $\bq\gtrsim0.1\,k_F$, a value easily reached in Raman spectroscopy experiments even with a small angle between the Raman beams, the shift is negligible. This establishes a major advantage of Raman spectroscopy over conventional rf spectroscopy.

In the inset of Fig.~\ref{fig:peak_position_vs_q}, we calculate the shift in the rf spectroscopy peak at unitarity as a function of temperature, at finite impurity density. The calculated rf shift initially decreases and then increases with temperature.
This non-monotonic dependence follows the trend of the initial-to-final state energy gap at the maximally-populated momentum value.
Interestingly, even without considering a finite-temperature reservoir, we find that for $T_p\gtrsim 0.1\,T_F$, the shift increases with temperature, as observed in Ref.~\cite{Yan2019} and discussed in Refs.~\cite{Tajima2018,Mulkerin2019,Tajima2019,Liu2020}.

\section{Raman transition rate of the coherent polaron contribution in the LDA}\label{sect:appendix_raman_coherent}
In this appendix we provide more detail on the derivation of the Raman rate for the coherent polaron contribution under the LDA.

Eq.\,\eqref{Eq_total_raman_rate_coh} gives the rate for a homogeneous system. Here, we treat the case of a harmonically trapped gas. The occupation averaged coherent response is given by
\begin{align}\label{Eq:AnsatzRamanCohLDA}
\bar{\Gamma}_{\text{coh}}\left(\omega\right)&= \frac{2\pi\Omega^2_e}{V}\int\mathrm{d}^3\mathbf{r} \sum_\veck \mathcal{A}_{\text{coh}}\left(\omega,\kv\right)  \\ 
&\times n_F\left[\epsilon_{\mathrm{pol}}(\kv)-\mu + \frac{m}{2}\omega_{\mathrm{ho}}^2 \left(1-\frac{\Ep^0}{ \varepsilon_F}\right)\mathbf{r}^2,T_p\right]\;, \nonumber
\end{align}
where $\mathcal{A}_\text{coh}(\omega,\kv)$ is given in Eq.\,\eqref{Eq_spectral_function_of_polarons}, and $\omega_{\mathrm{ho}}=\epsilon_F/(6 N)^{1/3}$ denotes the geometrically-averaged harmonic trapping frequency. In cylindrical coordinates the integral over $\kv$ within Eq.\,\eqref{Eq:AnsatzRamanCohLDA} decomposes into a two-dimensional integral of $ \kv_\perp$ over directions perpendicular to $\dpv$, and an integral over $ k_{\bar{q}}$ along the direction of $\dpv$. The condition imposed by the $\delta$-function within $\mathcal{A}_\mathrm{coh}$ is then given by
\begin{equation}\label{eq:deltaconditionCohPol}
\frac{k_{\bar{q}}^{2}+k_{\perp}^{2}}{2}\left(\frac{1}{m}-\frac{1}{m^\ast}\right)+\frac{k_{\bar{q}}\bar{q}}{m}-\omega-\Ep^{0}+\frac{\bar{q}^{2}}{2m}=0\;,
\end{equation}
and can be solved for $k_{\bar{q}}$. 

At low temperatures and for most interaction strengths, we find that ${k\left(1-m/m^\ast\right)\ll \bar{q}}$ for momenta which are not suppressed by the Fermi distribution. We thus neglect the first term proportional to $(1- m/m^\ast)$ in Eq.\,\eqref{eq:deltaconditionCohPol} when evaluating the $\delta$-function in Eq.\,\eqref{Eq:AnsatzRamanCohLDA}. Carrying out the integrations in Eq.\,\eqref{Eq:AnsatzRamanCohLDA} we then obtain that
\begin{equation}
\bar{\Gamma}_{\text{coh}}(\omega)=2\pi m\Omega_e^2N_\text{coh}\left.n_{\text{P}}\left[k_{\bq}\left(\omega\right)\right]\right/{\bar{q}}\;, 
\label{eq:Gamma_coh}
\end{equation}
where $n_P[k]$ is given in Eq.\,\eqref{Eq_momentum_distribution_of_minority} and $\kq(\omega)$ is given in Eq.\,\eqref{Eq_defining_kq_for_polarons}. Correspondingly, the fugacity $\zeta_{\text{P}}=e^{-\left.\left(\Ep^0-\mu\right)\right/T_p}$ within Eq.\,\eqref{Eq_momentum_distribution_of_minority} is set by the normalization $\int\mathrm{d}\omega\bar{\Gamma}_{\text{coh}} (\omega)= 2 \pi \Omega^2_e N_{\text{coh}}$, which gives
\begin{equation}\label{Eq_fugacity_of_minority} 
\mathrm{Li}_3\left(-\zeta_{2}\right)=-\frac{x\bar{Z}}{6}\left[\frac{\EF\left(\EF-\Ep^0\right)}{\frac{m^\ast}{m^{\phantom{\ast}}}T_p^2}\right]^{3/2}\;.
\end{equation}

\section{Raman transition rate of the incoherent polaronic and molecular contributions
\label{mol_spectrum}}

Here we derive the response of a thermal ensemble of molecules, each of which is made of a single impurity and a single bath particle and considered to be in vacuum. As a wavefunction Ansatz for the molecule, we use 
 \begin{align}
     |\psi^{\kv_{\text{cm}}}\rangle=\sum_{\lv}\gamma_{\lv}^{\kv_{\text{cm}}}c_{-\lv}^{\dagger}d_{\lv+\kv_{\text{cm}}}^{\dagger}|0\rangle\;.
     \label{Eq:Wavefunction_Vacuum_Molecule}
 \end{align}
Note that the only difference to Eq.\,\eqref{Eq:WaveFunctionMolecule} is that, here, we do not consider the Fermi sea of background particles. This simplification is made in order to obtain a closed-form expression for the fitting function which is feasible to calculate numerically. Using Eq.\,\eqref{Eq:Wavefunction_Vacuum_Molecule} as the initial state $|i\rangle$ for Eq.\,\eqref{Eq:FermisGoldenRule}, one obtains 
\begin{align}
    \mathcal{A}\big(\omega, \psi&^{\kv_{\text{cm}}}\big)= \\
    &\sum_{\lv} {}\left|\gamma_{\mathbf{l}}^{\kv_{\text{cm}}}\right|^{2} \delta\left(\omega-\varepsilon_{\mathbf{l}}-\varepsilon_{\kv_{\text{cm}}-\mathbf{l}+\overline{\mathbf{q}}}+E_b+\frac{k_{\text{cm}}^2}{4m}\right)\;.\nonumber
\end{align}
The variational parameter $\gamma_{\mathbf{l}}^{\mathbf{k}_{\text{cm}}}$ has to be obtained by the minimization of $\langle \psi^{\kv_{\text{cm}}}|\mathcal{H}-E|\psi^{\kv_{\text{cm}}}\rangle$.
Within this calculation the relative momentum after the Raman dissociation is given by $\kv_{\mathrm{rel}}= \lv + (\kv_{\text{cm}}+ \dpv)/2 $. After changing variables $\lv$ to $\kv_{\mathrm{rel}}$, evaluating the $\delta$-function implements the energy conservation of Eq.\,\eqref{Eq_k_rel_squared}. The probability $F(\kv_{\mathrm{rel}})$ of Eq.\,\eqref{Eq_k_rel_distribution} is then given directly by the matrix element $|\gamma_{\kv_{\mathrm{rel}}- (\kv_{\text{cm}}+ \dpv)/2}^{\mathbf{k}_{\text{cm}}}|^{2}$. Averaging the Raman spectral function over all momenta $\kv_{\text{cm}}$ weighted by a thermal Boltzmann distribution yields
\begin{equation}
   \bar{\mathcal{A}}(\omega)=\int\mathrm{d}\kv_{\text{cm}} \ \mathcal{A}\left(\omega, \psi^{\kv_{\text{cm}}}\right)\cdot G\left(\kv_{\text{cm}}, T_{\mathrm{bg}}\right)\;.
\end{equation}
The final expression in Eq.\,\eqref{Eq_P_cpairs_full} is obtained by subsequent integration of $\kv_{\text{cm}}$ over directions perpendicular to $\dpv$.

\section{Validation of the fitting model}
\label{app:fitting_val}

\begin{figure}
	\centering
	\includegraphics{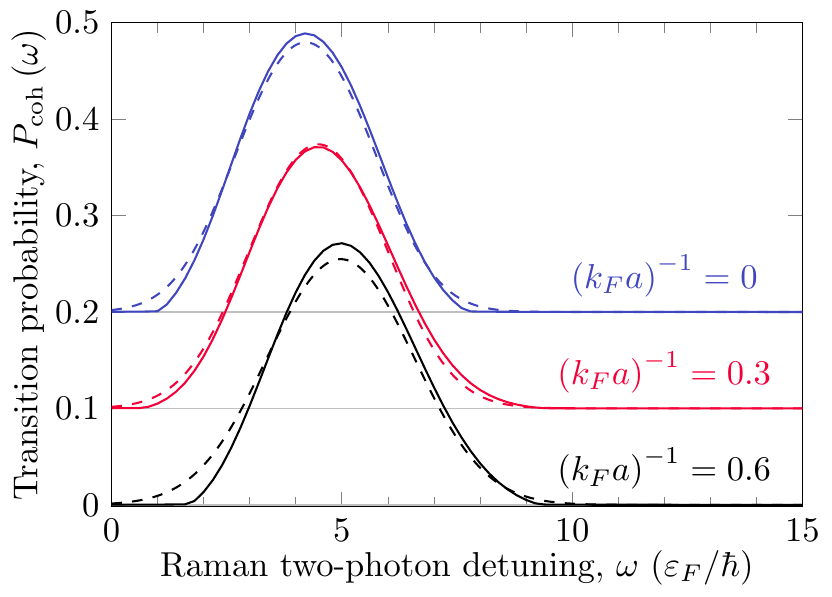}
	\caption{\textbf{Comparison of coherent part Raman spectra.} Using the full many-body model for the Raman spectra (first term of Eq.\,\eqref{Eq:RamanPolaronCohAndInc}, solid lines), and the approximation due to Eq.\,\eqref{EQ_trans_prob_QP} (dashed lines), the coherent Raman rate is shown for three interaction strengths. In these calculations we use the effective mass and polaron energy obtained from the polaron Ansatz. For clarity, the second and third graphs from the bottom are shifted by $0.1$ and $0.2$, respectively.}
	\label{fig:comparison_full_and_approximation_Raman_rate}
\end{figure}

In this appendix we discuss the applicability of our fitting model, i.e., Eqs.\,\eqref{EQ_trans_prob_QP} and \eqref{Eq_P_cpairs_full}, to the Raman spectra of the impurity problem. To this end, we compare the two parts of the fitting function to theoretical calculations. 

\textbf{Coherent polaron response.---} In Fig.~\ref{fig:comparison_full_and_approximation_Raman_rate} we present a comparison between the first part of the fitting model, namely $P_{\mathrm{coh}}$ (dashed lines), and the full solution of our theoretical model introduced in Section~\ref{sect:FermiPolaronModel} (solid lines). As can be seen, the approximation is excellent at unitarity and at $\kfainv=0.3$. Closer to the predicted transition, minor differences develop at $k_q\approx k_F$. There are two causes to this behavior. First, the increase of $m^\ast$ leads to a small asymmetry. Second, the polarons do not populate high momentum states since the width of the excitation branch increases dramatically as the momentum increases, leading to a narrowing of the theoretical Raman spectrum. Importantly, the center peak position coincides for both spectra, which allows us to use Eq.\,\eqref{Eq_omega_0_of_polaron} for the extraction of $\Ep^0$.

\begin{figure}
	\centering
	\includegraphics{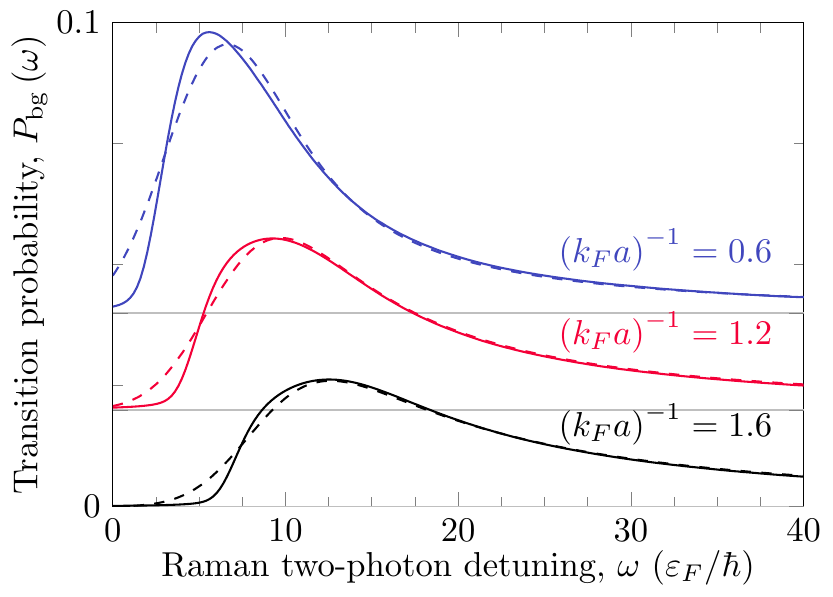}
	\caption{\textbf{Comparison of background Raman spectra.} Using the full many-body model for the Raman spectra (incoherent and molecular terms in Eq.\,\eqref{Eq:FullRamanSpectrum}, solid lines), and the approximation obtained by fitting the simplified model Eq.\,\eqref{EQ_trans_prob_QP} to the theoretical spectra (dashed lines), Raman spectra are shown for three interaction strengths. For clarity, the second and third graphs from the bottom are shifted by $0.02$ and $0.04$, respectively.}
	\label{fig:bg_mol_fit}	
\end{figure}

\textbf{Background signal.---} Here, we analyze the applicability of the second part of our fitting function, $P_{\mathrm{bg}}$, to fit the background spectrum that combines the incoherent and molecular contributions. In Fig.~\ref{fig:bg_mol_fit}, we compare the best fit of $P_{\mathrm{bg}}$ to the background signal, as calculated by the full many-body model. Overall they match well, especially at high frequencies. The difference at low frequencies stems from the neglect of the majority specie's Fermi surface in the fitting model.

We should consider systematic errors in extracted observables that may arise due to this approximation. $T_{\mathrm{bg}}$ affects almost solely the low-frequency part of the spectrum. Therefore, it should be chosen to compensate for the absence of Pauli blocking in our fitting model and minimize errors in the extracted $\bar{Z}$. The effective binding energy, $E_b$, on the other hand, affects mainly the high-frequency part of the spectrum, where the fit and numerical data are in excellent agreement.

\begin{figure}
	\centering
	\includegraphics{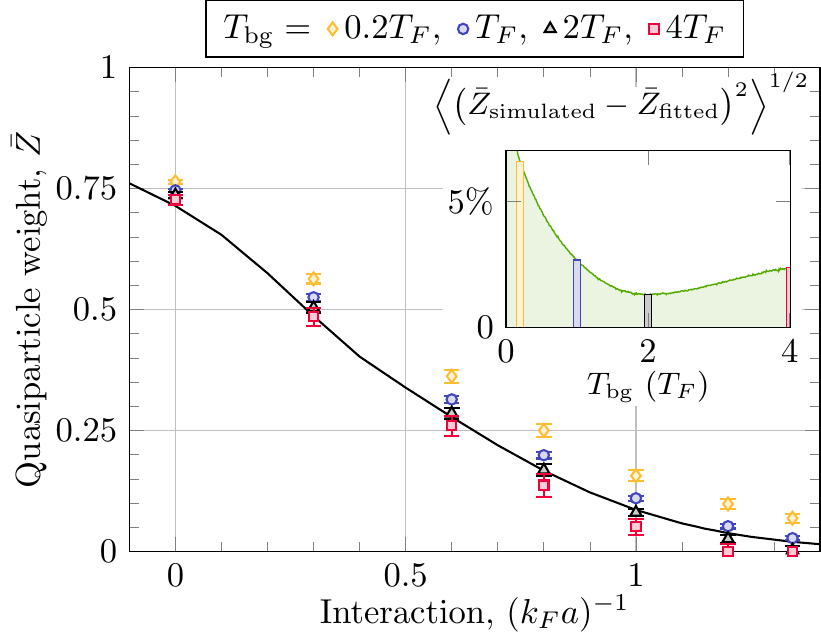}
	\caption{\textbf{Fitting simulated background Raman spectra with fixed effective temperature $T_{\mathrm{bg}}$.} The black line denotes the homogeneous many-body quasiparticle weight computed for $T=0.2\,T_F$, $x=0.15$. Errorbars mark the fitted residue using an effective temperature of $0.2\,T_F$ (yellow diamonds), $T_F$ (blue circles), $2\,T_F$ (black triangles), and $4\,T_F$ (red squares). Inset: Root-mean-square deviation of the extracted residue from the computed one, exhibiting a minimal deviation at approximately $2\,T_F$, the value we use for fitting the experimental data.}
	\label{fig:T_ncp_opt}	
\end{figure}

We find the optimal value for $T_\mathrm{bg}$ by fitting theoretical Raman spectra of the background signal (incoherent polaron and molecule) due to Eq.\,\eqref{Eq:FullRamanSpectrum} at eight interaction strengths. In Fig.~\ref{fig:T_ncp_opt}, we plot the fit results for the quasiparticle residue, obtained with four exemplifying values of $T_\mathrm{bg}$. The effect of varying $T_\mathrm{bg}$ is a systematic shift of the residue. The inset of Fig.~\ref{fig:T_ncp_opt} presents the root-mean-square difference between the simulated and the fitted quasiparticle residue as a function of the fixed value for the effective temperature. We observe a minimal discrepancy at $T_\mathrm{bg}\approx2\,T_F$. Fig.~\ref{fig:T_ncp_opt} clearly shows that even at sub-optimal values of $T_\mathrm{bg}$, the qualitative behavior of $\bar{Z}$ does not change. The reason for this is that $\bar{Z}$ measures the spectral weight of the roughly symmetric peak, and therefore it is rather insensitive to variations in the fitting procedure.

\section{Setting the chemical potential}
\label{sect_app_chempots}
\begin{figure}
	\centering
	\includegraphics{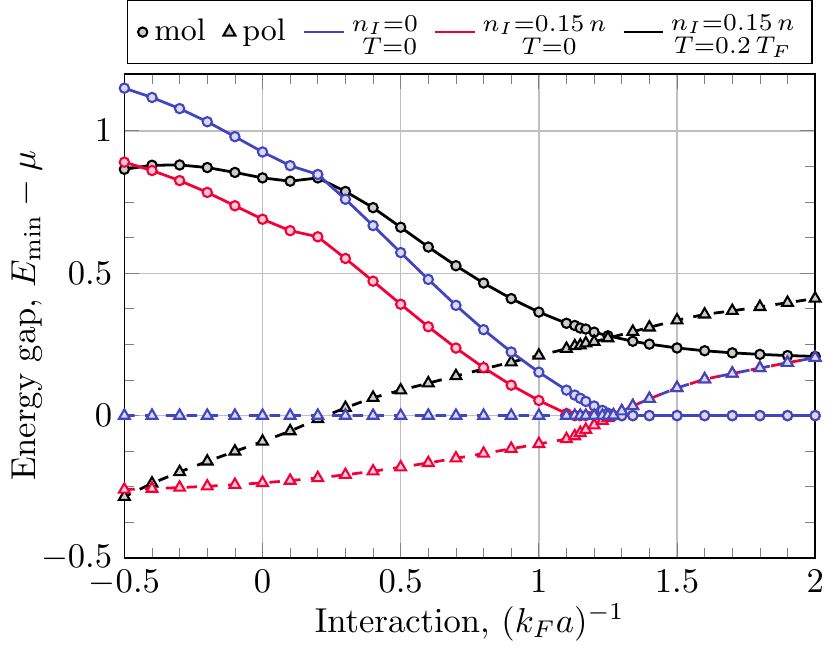}
\caption{\textbf{Polaronic and molecular energy gap.} Energy gap between the lowest-lying polaron/molecule modes and the chemical potential vs. interaction strength. The blue lines show the gap for the case of a single impurity at $T=0$, while the red and black lines show the gaps for finite density at $T=0$ and $T>0$, respectively. The solid lines with circle markers denote the gap between the lowest-lying molecular state and the chemical potential, whereas the dashed lines with triangle markers denote the polaronic gaps.
}
\label{Fig:ChemicalPotentialTheory}
\end{figure}

In this appendix we elaborate on the behavior of the chemical potential with respect to the polaron and molecule dispersions, intending to highlight the mechanism by which the bands are populated. Throughout this manuscript, we tune the chemical potentials such that under application of Eq.\,\eqref{Eq:ImpDensity} the respective densities are reproduced, which means that the polaron and the molecule are populated according to fermionic and bosonic statistics, respectively.

Tuning the chemical potential $\mu\to-\infty$ yields a vanishing density. As the chemical potential is increased, the two bands begin to be populated accordingly such that a small impurity density begins to form. Tuning the chemical potential further, eventually it will reach the minimum of the lower-lying dispersion. If the polaron dispersion is lower-lying, it may surpass the minimum, effectively forming a Fermi surface. Tuning even further, the chemical potential will reach the minimum of the molecule dispersion. As we populate according to bosonic statistics, at sufficiently low temperature, the molecules will thus begin to condense in the corresponding state and a molecular BEC will form within the fermionic bath. The chemical potential may therefore not surpass this minimum and will remain pinned to it even for larger impurity densities.

In Figure~\ref{Fig:ChemicalPotentialTheory} the energy gap between the chemical potential and the lowest-lying polaronic (triangles, dashed) and molecular (circles, solid) states is shown for different interaction strengths. We show this gap for the case of a single impurity (blue), at finite density ($n_I=0.15\,n$) and $T=0$ (red) as well as at finite density and finite temperature ($T=0.2\,T_F$, $n_I=0.15\,n$, black). In the single impurity limit the, the chemical potential is tuned to the minimum of $\Ep(\pv)$ below the transition and to the minimum of $\Em(\pv)$ above. Thus, below the transition the gap of the molecular dispersion marks the energy gap between the polaronic ground state and the excited molecular state, and vice versa above the transition. 

At finite density and zero temperature, below the transition a Fermi surface of polarons forms. When the chemical potential approaches the lowest-lying molecular state (solid, red) at around $\kfainv\approx 1.1$ molecules begin to condense. Beyond the transition, the polaronic Fermi surface vanishes (positive energy gap) and all impurities condense in the molecular ground-state. At finite temperature, a polaronic Fermi surface forms initially ($\kfainv\lesssim 0.2 $), but eventually the polaron and the molecule are both populated thermally, as visible from their positive energy gaps. Note that the lowest point of the dispersion relations does not necessarily lie at $\pv=0$.

\section{Single-impurity contact coefficients}\label{Appendix:SingleImpContact}
Here we provide generalized expressions for the single-impurity contact at finite momentum, as derived for ${\pv=0}$ in Refs.~\cite{Punk2009,punk2010thesis}. 
They read
\begin{align}
&C_{\mathrm{pol}}\left[\pv,\Ep(\pv)\right]= \\&\qquad \frac{1}{V}
\sum_{\mathbf{q}}{}^{'}\frac{m^2\left|\alpha_0^{\pv}\right|^{2}}{\left|\frac{1}{U}-\frac{1}{V} \sum_{\kv}{}^{'} \frac{1}{\Ep(\pv)-\varepsilon_{\mathbf{k}}-\varepsilon_{\mathbf{q}-\mathbf{k}+\pv}+\varepsilon_{\mathbf{q}}}\right|^2}\nonumber
\end{align}
for the polaron, and
\begin{align}
&C_{\mathrm{mol}}\left[\pv,\Em(\pv)\right]=\\&\qquad
m^2 \left[\frac{1}{V}\sum_{\kv}{}^{'} \left|\frac{1}{\Em(\pv)+\varepsilon_{F}-\varepsilon_{\mathbf{k}}-\varepsilon_{\pv+\kv}}\right|^2\right]^{-1} \nonumber
\end{align}
for the molecule.
From this, the full many-body contact coefficient $C$ can be calculated via Eq.\,\eqref{Eq:ManyBodyContact}.

\end{document}